\documentclass[11pt,a4paper]{article}
\usepackage{amsmath}
\usepackage{amsfonts}
\usepackage{color}
\usepackage{graphicx}
\usepackage{float}
\usepackage{subfigure}
\graphicspath{{figures/}}
\usepackage{dcolumn}
\usepackage{bm}
\usepackage[
bookmarks,bookmarksnumbered,colorlinks=true,anchorcolor=blue,
linkcolor=blue,urlcolor=blue,citecolor=blue,
breaklinks=true]{hyperref}
\usepackage{geometry}
\usepackage{hep}
\usepackage{extarrows}
\usepackage{authblk}
\usepackage{amssymb}
\usepackage[numbers,sort&compress]{natbib}
\usepackage{cite}
\usepackage{pifont}
\usepackage{tikz}
\usepackage{sansmath}
\usepackage{mathrsfs}
\usetikzlibrary{shadings,intersections,snakes}
\usepackage{verbatim}
\numberwithin{equation}{section}
\usepackage{tikz}
\usetikzlibrary{calc}

\date{\today}

\def\a{{\rm a}}

\def\i{{\rm i}}

\begin{document}

\title{\bf Chiral phase transition: effective field theory and holography}

\author[]{Yanyan Bu \thanks{yybu@hit.edu.cn}~}

\author[]{Zexin Yang \thanks{24b311002@stu.hit.edu.cn}~}

\affil[]{\it School of Physics, Harbin Institute of Technology, Harbin 150001, China}

\maketitle

\begin{abstract}
We consider the chiral phase transition relevant for QCD matter at finite temperature but with vanishing baryon density. Presumably, the chiral phase transition is of second order for two-flavor QCD in the chiral limit {\color{red}\cite{Cuteri:2021ikv}}. Near the transition temperature, we apply the Schwinger-Keldysh formalism and construct a low-energy effective field theory for the system, in which fluctuations and dissipations are systematically captured. The dynamical variables involve the chiral charge densities and order parameter (chiral condensate). Via the holographic Schwinger-Keldysh technique, the effective action is further confirmed within a modified AdS/QCD model. With higher-order terms suitably neglected, the stochastic equations derived from the effective field theory resemble those of model F in the Hohenberg-Halperin classification. Within the effective field theory, we briefly discuss the spontaneous breaking of chiral symmetry and Goldstone modes.
\end{abstract}

{\let\thefootnote\relax\footnotetext{Authors are ordered alphabetically and should be both considered as co-first authors as well as co-corresponding authors.} }

\newpage

\tableofcontents

\allowdisplaybreaks

\flushbottom

\section{Introduction}

Color confinement and chiral symmetry breaking ($\chi$SB) are two important features of non-perturbative Quantum Chromodynamics (QCD), both playing crucial roles in our understanding of the strong interactions. The mechanism of the color confinement has been a long-standing problem and remains mysterious. Nevertheless, the color confinement implies that in the low-energy regime of QCD in vacuum, the degrees of freedom are no longer quarks and gluons, but rather hadrons. The $\chi$SB has been extensively studied for QCD in vacuum, with chiral perturbation theory established as a low-energy effective framework for describing the dynamics of light hadrons (e.g., pions), which are light excitations around the QCD vacuum.

The properties of QCD matter at finite temperature $T$, baryon chemical potential $\mu_{\rm B}$, etc have been an active research topic for many years \cite{Jaiswal:2020hvk,MUSES:2023hyz}. Currently, thanks to both theoretical and experimental efforts, a phase diagram for QCD matter has been conjectured over a broad range in the $(T,\mu_{\rm B})$-plane \cite{Halasz:1998qr,Bzdak:2019pkr,Du:2024wjm}. The commonly conjectured phase diagram predicts some interesting phases of QCD matter, which might be of relevance in laboratories or compact stars \cite{MUSES:2023hyz}. Indeed, with the help of high energy accelerators 
(e.g., the ongoing Beam Energy Scan program in the Relativistic Heavy Ion Collider (RHIC) \cite{Luo:2020pef} and future facilities like ``Facility for Antiproton and Ion Research (FAIR)'' and ``Nuclotron-based Ion Collider fAcility (NICA)'' \cite{Senger:2021cfo})
, experimental physicists are dedicated to searching for potential signals (dis)favoring the conjectured QCD phase diagram. A particular focus is on the existence of a critical endpoint and, if existed, its location on the $(T, \mu_{\rm B})$-plane.

QCD matter contains more fruitful physics than its vacuum counterpart. Meanwhile, the dynamics of QCD matter is inevitably much more involved, partially due to medium effects, complicated many-body dynamics, etc. This fact motivated one to pursue new ideas or even new methodologies for studying QCD matter under various conditions such as finite $T$ and finite $\mu_{\rm B}$. Among others, effective models were proposed to explore the properties of QCD matter in a certain window. From the symmetry perspective, QCD near the chiral phase transition is effectively described by the O(4) model G \cite{Pisarski:1983ms,Rajagopal:1992qz} of the Hohenberg-Halperin classification \cite{RevModPhys.49.435}, since the two-flavor QCD in the chiral limit has an exact chiral symmetry SU(2)$_L$ $\times$ SU(2)$_R$ $\simeq $ O(4). It is thus tempting to study the dynamics of QCD near the chiral phase transition by adopting ideas from the theory of dynamic critical phenomena \cite{RevModPhys.49.435}, e.g., the real-time pion propagation problem resolved in \cite{Son:2002ci,Son:2001ff}. On the other hand, in the long-wavelength long-time regime, an ideal hydrodynamic framework was formulated \cite{Son:1999pa} for QCD matter in the chiral limit. Within such a framework, the dynamic variables reflect not only conserved quantities but also the pions arising from spontaneous $\chi$SB. In recent years, these effective approaches have been further refined to address fluctuation contributions \cite{Grossi:2020ezz,Grossi:2021gqi,Florio:2021jlx,Cao:2022csq,
Braun:2023qak,Roth:2024rbi} to dynamical quantities like transport coefficients for QCD matter near the chiral phase transition.

In the past decade, by virtue of the Schwinger-Keldysh (SK) formalism, dissipative hydrodynamics has been reformulated as a Wilsonian effective field theory (EFT) \cite{Crossley:2015evo,Glorioso:2017fpd,Haehl:2015uoc,Haehl:2018lcu} (see \cite{Liu:2018lec} for a review), which will be referred to as hydrodynamic EFT below. The hydrodynamic EFT provides a promising framework for studying the real-time dynamics of out-of-equilibrium QCD matter, particularly in the systematic treatment of fluctuations and dissipations. Indeed, the formulation of hydrodynamic EFT has been largely enlightened by holographic duality \cite{Maldacena:1997re,Gubser:1998bc,Witten:1998qj}. Moreover, holographic prescriptions for the SK formalism \cite{Herzog:2002pc,Skenderis:2008dh,Skenderis:2008dg,Glorioso:2018mmw} make it possible to {\it derive} boundary effective action for a certain bulk theory, see, e.g., \cite{Glorioso:2018mmw,deBoer:2018qqm,Chakrabarty:2019aeu,Bu:2020jfo,Bu:2021clf,
Bu:2021jlp,Bu:2022esd,Bu:2022oty,Baggioli:2023tlc,Bu:2024oyz,Liu:2024tqe,Baggioli:2024zfq}\footnote{Similar studies were carried out in \cite{Ho:2013rra,Ghosh:2020lel,He:2021jna}. We understand that it is on the Wilsonian influence functional rather than on the off-shell effective action that was focused therein.} for recent developments. Holographic derivation of hydrodynamic EFT is of importance on its own right: (I) it would help to understand/examine postulated symmetries that are pivotal in formulating hydrodynamic EFT, and may even shed light on possible extension of current hydrodynamic EFT; (II) it provides knowledge for parameters in an EFT whose underlying microscopic theory involves a strongly coupled quantum field theory.

The present work aims at formulating a SK EFT for two-flavor QCD near the chiral phase transition. The goal is achieved through two complementary approaches: the hydrodynamic EFT of \cite{Crossley:2015evo,Glorioso:2017fpd,Liu:2018lec} versus the holographic SK technique of \cite{Glorioso:2018mmw}. 
Notice that the model G concerns the dynamics of the chiral charges and the chiral condensate \cite{Pisarski:1983ms,Rajagopal:1992qz}; particularly, the energy and momentum are frozen. Thus, we will ignore the variations of energy and momentum densities throughout this study. Nevertheless, it should be pointed out that the inclusion of energy and momentum dynamics would render the universality class of real-world QCD to be that of model H \cite{Son:2004iv}.
Recently, a hydrodynamic EFT has been constructed for conserved charges associated with an internal non-Abelian symmetry \cite{Glorioso:2020loc,Bu:2022esd,Hongo:2024brb}. In this work, we extend the construction of \cite{Glorioso:2020loc,Bu:2022esd,Hongo:2024brb} by adding a non-Abelian SU(2)$_L$ $\times$ SU(2)$_R$ scalar, which is the fluctuating chiral condensate of a two-flavor QCD. This is mainly motivated by the critical slowing-down phenomenon, which indicates that the non-conserved chiral condensate evolves slowly near the phase transition. Therefore, in addition to the conserved charges, the chiral condensate shall be retained as a dynamical variable in the low-energy EFT.

The EFT to be presented below stands for a non-Abelian generalization of that for a nearly critical U(1) superfluid \cite{Donos:2023ibv,Bu:2024oyz}. However, this does not necessarily mean our study will be bland or straightforward. On the one hand, the non-Abelian symmetry renders both the EFT construction and the derivation of stochastic equations rather non-trivial. On the other hand, we shall carefully tune the mass of a bulk scalar field so that the dual system is around the critical point and the dual operator has the desired dimension of a chiral condensate. This inevitably makes the holographic calculations very challenging, particularly on searching for solutions to the bulk scalar field.


Here, we connect the present study with previous works \cite{Son:2004iv,Natsuume:2010bs} and clarify some confusions. The work \cite{Son:2004iv} considered the real-world QCD so that the dynamical variables in the hydrodynamic limit are associated with the conserved densities (the energy, momentum, and baryon number) and the chiral condensate (indeed, merely the amplitude of the order parameter). It turns out that the baryon number density and the chiral condensate get mixed, leaving out only one truly hydrodynamic mode---a linear combination of the two. Thus, the actual dynamic universality class of QCD was claimed to be that of the model H. In \cite{Natsuume:2010bs} the authors considered the SU($N_c$) super-Yang-Mills theory with a global U(1) symmetry, mimicking the QCD. However, in the large $N_c$ limit, the mode coupling effect is suppressed, rendering the theory to belong to the model B. So, the different conclusions on the QCD universality class is mainly due to the different scenarios undertaken.

Admittedly, in order to meet the actual scenario investigated in the Beam Energy Scans in the heavy-ion collisions, the present work shall be extended in a number of aspects. First, the flavor symmetry will be enlarged to the U(3)$_L\times$ U(3)$_R$, where the vectorial part U$_V$(3) will be non-anomalous. Second, a matrix-valued source $M$ will be turned on for the chiral condensate, reflecting finite masses for $u$-, $d$-, $s$-quarks. Then, a term like $M^{\dagger} \Sigma + M \Sigma^{\dagger}$ will be present in the EFT action. Ignoring the mass difference between $u$ and $d$, the vectorial part of flavor symmetry will be explicitly broken to U$_B$(1)$\times$SU$_I$(2)$\times$U$_S$(1), which correspond to the conserved baryon number, isospin, and strangeness, respectively. Third, the dynamics of energy and momentum will be considered and their coupling to the flavor sector is important in determining QCD universality class \cite{Son:2004iv}. The construction of SK EFT for such a more complete scenario will be left as a future study. The holographic setup may be extended in parallel, see discussion in section \ref{holo_study}.

The rest of this paper will be organized as follows. In section \ref{EFT_constru}, we present the EFT construction. We also recast the EFT into a stochastic formalism and compare it with the model F of \cite{RevModPhys.49.435}. In section \ref{holo_study}, we present a holographic derivation of the EFT constructed in section \ref{EFT_constru}. Here, we consider an improved AdS/QCD model \cite{Chelabi:2015gpc,Chelabi:2015cwn,Chen:2018msc,Cao:2022csq}, which realized $\chi$SB spontaneously by modifying the mass of a bulk scalar field in the AdS/QCD model \cite{Erlich:2005qh}. In section \ref{sum_and_outl}, we provide a brief summary and outlook some future directions.

\section{Effective field theory for chiral phase transition}\label{EFT_constru}

In this section, we consider the chiral phase transition for two-flavor QCD at finite temperature and zero baryon density. We will focus on the chiral limit so that there is an exact SU(2)$_L$ $\times$ SU(2)$_R$ flavor symmetry. In addition, we assume that the  temperature is slightly above a critical value at which the chiral phase transition happens. This means that the flavor symmetry is not spontaneously broken, which will simplify the study. In the long-wavelength long-time limit, we search for a low-energy EFT description for such a system. The dynamical variables contain both the conserved charges associated with the flavor symmetries and the non-conserved order parameter characterizing the chiral phase transition.

\subsection{Dynamical variables and symmetries} \label{variable_symmetry}

The flavor symmetry implies conserved chiral currents $J_L^\mu$ and $J_R^\mu$
\begin{align}
\partial_\mu J_L^\mu = 0, \qquad \qquad \partial_\mu J^\mu_R =0 \label{conservation}
\end{align}
The conservation laws \eqref{conservation} can be ensured by coupling the currents $J_L^\mu$ and $J_R^\mu$ to background gauge fields $\mathcal A_\mu$ and $\mathcal V_\mu$ respectively, and further requiring the theory to be invariant under the gauge transformations of the background gauge fields
\begin{equation}
\mathcal{A}_\mu \to e^{\i \lambda^{\a}(x) t^{\a}} \left(\mathcal{A}_\mu + \i \partial_{\mu}\right) e^{-\i \lambda^{\a} t^{\a}}, \qquad \qquad  \mathcal{V}_\mu \to e^{\i \chi^{\a}(x) t^{\a}} \left(\mathcal{V}_\mu + \i \partial_{\mu}\right) e^{-\i \chi^{\a}(x) t^{\a}} \label{non-dynamical_gauge_transform}
\end{equation}
Here $\lambda^\a(x)$ and $\chi^\a(x)$ are arbitrary functions generating the non-dynamical gauge transformations, and $t^{\a} = \sigma^\a/2$ are the SU(2) generator with $\sigma^\a$ the Pauli matrices. Meanwhile, we have the order parameter $\mathcal O$ transforming as a bi-fundamental scalar
\begin{align}
\mathcal O \to e^{\i \lambda^\a(x) t^\a} \mathcal O e^{-\i \chi^\a(x) t^\a} \label{O_gauge_transform}
\end{align}
The low-energy EFT is demanded to preserve the non-dynamical gauge symmetry \eqref{non-dynamical_gauge_transform}-\eqref{O_gauge_transform} so that \eqref{conservation} are automatically satisfied.

This idea motivates us to promote the gauge transformation parameters $\lambda^\a(x)$ and $\chi^\a(x)$ to dynamical fields and identify them as the suitable dynamical variables in the EFT \cite{Crossley:2015evo}. Immediately, we are led to the following combinations
\begin{align}
B_\mu \equiv \mathcal U(\varphi) \left(\mathcal{A}_\mu + \i \partial_{\mu}\right) \mathcal U^{\dagger}(\varphi), \qquad \qquad C_\mu \equiv \mathcal U(\phi) \left(\mathcal{V}_\mu + \i \partial_{\mu}\right) \mathcal U^{\dagger}(\phi)  \label{B_C}
\end{align}
where
\begin{align}
\mathcal U(\varphi) = e^{\i \varphi^\a(x) t^\a}, \qquad \qquad \mathcal U (\phi) = e^{\i \phi^\a(x) t^\a}  \label{U_LR}
\end{align}
Accordingly, instead of $\mathcal O$, it will be more convenient to work with
\begin{align}
\Sigma \equiv \mathcal U(\varphi) \mathcal O \mathcal U^\dagger (\phi) \label{Sigma}
\end{align}
Note that $B_\mu$, $C_\mu$ and $\Sigma$ are invariant under the non-dynamical gauge transformations \eqref{non-dynamical_gauge_transform} and \eqref{O_gauge_transform} if $\varphi$ and $\phi$ also participate in this non-dynamical gauge transformation via shifts
\begin{align}
\varphi^\a \to \varphi^\a - \lambda^\a, \qquad  \phi^\a \to \phi^\a - \chi^\a
\end{align}
Therefore, $B_\mu$, $C_\mu$ and $\Sigma$ are the ideal building blocks for constructing the EFT action. Notice that in the EFT, $\varphi$, $\phi$ and $\Sigma$ are the dynamical fields whereas $\mathcal A_\mu$ and $\mathcal V_\mu$ act as external sources for the conserved chiral currents $J_L^\mu$ and $J_R^\mu$.

In the spirit of the SK formalism, all dynamical variables and external sources are doubled
\begin{align}
&\varphi \to \varphi_1, \varphi_2, \qquad \phi \to \phi_1, \phi_2, \qquad \mathcal A_\mu \to \mathcal A_{1\mu}, \mathcal A_{2\mu}, \qquad \mathcal V_\mu \to \mathcal V_{1\mu}, \mathcal V_{2\mu} \nonumber \\
&B_\mu \to B_{1\mu}, B_{2\mu}, \qquad C_\mu \to C_{1\mu}, C_{2\mu}, \qquad \Sigma \to \Sigma_1, \Sigma_2 \label{SK_doubling}
\end{align}
In the Keldysh basis, we have
\begin{align}
B_{r\mu} \equiv \frac{1}{2} (B_{1\mu} + B_{2\mu}), \qquad B_{a\mu} \equiv B_{1\mu} - B_{2\mu} \label{Bmu_Keldysh}
\end{align}
and similarly for the other variables.

The partition function of the system would be expressed as a path integral over the low-energy dynamical variables
\begin{align}
Z = \int [D \varphi_r] [D \varphi_a] [D \phi_r][D \phi_a][D \Sigma_r][D \Sigma_a] e^{\i S_{eff}[B_{r\mu}, C_{r\mu}, \Sigma_r; B_{a\mu}, C_{a\mu}, \Sigma_a ]} \label{Z_EFT}
\end{align}
where $S_{eff}$ is the EFT action. The action $S_{eff}$ is constrained by various symmetries which we briefly summarize here. Further details regarding these symmetries can be found in \cite{Crossley:2015evo,Glorioso:2017fpd,Liu:2018lec}.

(1) The constraints implied by the unitarity of time evolution
\begin{align}
& S_{eff}[\mathcal X_r; \mathcal X_a = 0] = 0, \label{normalization} \\
& \left( S_{eff}[\mathcal X_r; \mathcal X_a] \right)^* = - S_{eff}[\mathcal X_r; - \mathcal X_a], \label{Z2SYM} \\
& {\rm Im} \left( S_{eff} \right) \geq 0, \label{positive_condition}
\end{align}
where $\mathcal X$ collectively denotes $B_\mu$, $C_\mu$ and $\Sigma$.

(2) Spatially rotational symmetry. This guides one to classify the building blocks and their derivatives according to the spatially rotational transformation.

(3) Flavor SU(2)$_L$ $\times$ SU(2)$_R$ symmetry. Physically, this symmetry governs the coupling between the chiral charge densities and complex order parameter. In the high-temperature phase, the flavor symmetry SU(2)$_L$ $\times$ SU(2)$_R$ is unbroken. Along with SK doubling \eqref{SK_doubling}, we have a doubled symmetry (SU(2)$_{L,1}$ $\times$ SU(2)$_{L,2}$) $\times$ (SU(2)$_{R,1}$ $\times$ SU(2)$_{R,2}$). However, it is the diagonal part (with respect to the SK double copy) of the doubled symmetry, denoted as SU(2)$_{L,\, diag}$ $\times$ SU(2)$_{R,\, diag}$, that the action $S_{eff}$ shall satisfy. This will be automatically obeyed once $\Sigma$ and $\Sigma^\dagger$ appear simultaneously in each term of the action $S_{eff}$.

(4) Chemical shift symmetry. The EFT action $S_{eff}$ is invariant under diagonal time-independent shift
\begin{align}
\varphi_r^\a \to \varphi_r^\a + \sigma_L^\a(\vec x), \qquad  \phi_r^\a \to \phi_r^\a + \sigma_R^\a(\vec x), \qquad {\rm others~unchanged}  \label{chemical_shift}
\end{align}
where $\varphi_r^\a$ and $\phi_r^\a$ shall be understood similarly as the definition \eqref{Bmu_Keldysh}. Physically, this symmetry arises from the fact that the flavor symmetry SU(2)$_L$ $\times$ SU(2)$_R$ is not broken spontaneously in the high temperature phase. Under the shift \eqref{chemical_shift}, various building blocks transform as
\begin{align}
&\Sigma_{a}\rightarrow \mathcal{L}\Sigma_{a}\mathcal{R}^{\dagger},\qquad \quad \Sigma_{r}\rightarrow \mathcal{L}\Sigma_{r}\mathcal{R}^{\dagger} \nonumber \\
&B_{r0}\rightarrow \mathcal{L}B_{r0}\mathcal{L}^{\dagger}, \qquad B_{a\mu}\rightarrow \mathcal{L}B_{a\mu}\mathcal{L}^{\dagger}, \qquad B_{ri}\rightarrow \mathcal{L}(B_{ri}+\text{i}\partial_i)\mathcal{L}^{\dagger} \nonumber \\
&C_{r0}\rightarrow \mathcal{R}C_{r0}\mathcal{R}^{\dagger}, \qquad C_{a\mu}\rightarrow \mathcal{R}C_{a\mu}\mathcal{R}^{\dagger}, \qquad C_{ri}\rightarrow \mathcal{R}(C_{ri}+\text{i}\partial_i)\mathcal{R}^{\dagger} \label{transformforBCSigma}
\end{align}
where $\mathcal{L} = e^{\i\sigma_{L}^{\mathrm{a}}(\vec{x})t^{\mathrm{a}}}$ and $\mathcal{R} = e^{\i\sigma_{R}^{\mathrm{a}}(\vec{x})t^{\mathrm{a}}} $ are the elements of SU(2) group that depend arbitrarily on space but are time-independent. Apparently, $\Sigma_{r, a}$ transform as bi-fundamental, $B_{r0}$, $B_{a\mu}$, $C_{r0}$ and $C_{a\mu}$ transform in the adjoint, while $B_{ri}$ and $C_{ri}$ transform as gauge connections. This observation guides us to define three covariant derivative operators
\begin{align}
\mathcal{D}_{Li} \equiv \partial_{i}-\text{i}[B_{ri},\cdot], \qquad \mathcal{D}_{Ri} \equiv \partial_{i}-\text{i}[C_{ri},\cdot], \qquad
\mathcal{D}_{i}=\partial_i-\text{i}B_{ri}\cdot+\text{i}\cdot C_{ri} \label{covariant_derivative_spatial}
\end{align}
It should be understood that $\mathcal{D}_{Li}$ will act on the left-handed fields $B_{r 0}$ and $B_{a\mu}$; $\mathcal{D}_{Ri}$ will act on the right-handed fields $C_{r 0}$ and $C_{a\mu}$; while $\mathcal{D}_{i}$ will act on the order parameter $\Sigma_{r,a}$.

The chemical shift symmetry \eqref{chemical_shift} sets stringent constraints on the action. First, $B_{ri}$ and $C_{ri}$ will appear in the action by three ways: via their time derivatives, through their field strengths such as $(\mathcal F_L)_{rij}\equiv \partial_i B_{rj}-\partial_{j} B_{ri}-\i [B_{ri},B_{rj}]$, or via covariant derivatives \eqref{covariant_derivative_spatial}. Second, all the rest fields appear in the action through the following three ways: by themselves, by their time derivatives or by covariant spatial derivatives with the help of \eqref{covariant_derivative_spatial}.

(5) Dynamical Kubo-Martin-Schwinger (KMS) symmetry. In the classical statistical limit, this symmetry is realized as \cite{Glorioso:2017fpd,Liu:2018lec}
\begin{equation}
S_{eff}[B_{r\mu}, C_{r\mu}, \Sigma_r; B_{a\mu}, C_{a\mu}, \Sigma_a] = S_{eff}[\hat B_{r\mu}, \hat C_{r\mu}, {\hat \Sigma}_r; \hat B_{a\mu}, \hat C_{a\mu}, \hat{\Sigma}_a] \label{KMSsymmetry1}
\end{equation}
where
\begin{align}
&\hat{B}_{r\mu}(-x)= B_{r\mu}(x), \qquad \,\,\, \hat{B}_{a\mu}(-x) = \left[B_{a\mu}(x) + \i \beta \partial_{0} B_{r\mu}(x)\right], \nonumber \\
&\hat{C}_{r\mu}(-x) = C_{r\mu}(x), \qquad \,\,\, \hat{C}_{a\mu}(-x) = \left[C_{a\mu}(x) + \i \beta \partial_{0} C_{r\mu}(x) \right], \nonumber \\
&\hat{\Sigma}_{r}(-x) = - \Sigma_{r}^{\dagger}(x), \qquad \quad \hat{\Sigma}_{a}(-x) = - \left[\Sigma_{a}^{\dagger}(x) + i \beta \partial_{0} \Sigma_{r}^{\dagger}(x)\right] \label{KMSsymmetry2}
\end{align}
Here, $\beta$ is the inverse temperature.

(6) Onsager relations. This requirement follows from the symmetry properties of the retarded (or advanced) correlation functions under a change of the ordering of operators \cite{Crossley:2015evo}. While for some simple cases, Onsager relations are satisfied automatically once dynamical KMS symmetry is imposed, this is not generically true \cite{Baggioli:2023tlc,Bu:2024oyz}.

\subsection{The EFT action}\label{EFT_action}

With the dynamical variables suitably parameterized and the set of symmetries completely identified, we are ready to write down the effective action. Basically, as in any EFT, we will organize the effective action by the number of fields and spacetime derivatives. Schematically, the effective action is split as
\begin{equation}
S_{eff} = \int{d^4x}\, {\rm Tr}\left( \mathcal{L}_{eff} \right) = \int{d^4x}\, {\rm Tr} \left( \mathcal{L}_{diff} + \mathcal{L}_{\Sigma} + \mathcal{L}_{3} + \mathcal{L}_{4} \right),
\end{equation}
where $\mathcal{L}_{diff}$ is the diffusive Lagrangian for the conserved chiral charges; $\mathcal{L}_{\Sigma}$ is for the order parameter; $\mathcal{L}_{3}$ and $\mathcal{L}_{4}$ stand for cubic and quartic interactions respectively. Throughout this work, we will be limited to the level of Gaussian white noises. This means that the Lagrangian will not cover terms having more than two $a$-variables. Moreover, we will neglect multiplicative noises\footnote{The only exception is for $\varpi_2$- and $\varpi_4$-terms in \eqref{L_BCSIGMA_3}, which originate from the KMS symmetry and inevitably generate multiplicative noises.}.

For the diffusive Lagrangian $\mathcal{L}_{diff}$, we truncate it to quadratic order in the diffusive fields $B_\mu, C_\mu$ and to second order in spacetime derivatives\footnote{While the spatial derivatives in \eqref{L_diff} do generate cubic terms, these terms are fully dictated by the chemical shift symmetry \eqref{chemical_shift}.}. The result is
\begin{align}
\mathcal{L}_{diff} & = a_0 B_{a0}B_{r0} + a_1 B_{a0} \partial_0 B_{r0} + a_2 B_{ai} \partial_0 B_{ri} + a_{3} B_{a0} \mathcal{D}_{Li}\left( \partial_0 B_{ri} \right) \nonumber \\
& + a_{4} B_{ai} \mathcal{D}_{Li} \left(\partial_0 B_{r0} \right) + a_{5} B_{a0} \mathcal{D}_{Li} \left(\mathcal{D}_{Li} B_{r0} \right) + a_6 (\mathcal F_L)_{rij} \mathcal D_{Li} B_{aj} \nonumber \\
& + a_{7} C_{a0} C_{r0} + a_{8} C_{a0} \partial_0 C_{r0}  + a_{9} C_{ai} \partial_0 C_{ri} + a_{10} C_{a0} \mathcal{D}_{Ri} \left(\partial_0 C_{ri} \right) \nonumber \\
& + a_{11} C_{ai} \left( \mathcal{D}_{Ri} \partial_0 C_{r0} \right) + a_{12} C_{a0} \mathcal{D}_{Ri} \left(\mathcal{D}_{Ri} C_{r0} \right) + a_{13}(\mathcal F_R)_{rij} \mathcal D_{Ri} C_{aj} \nonumber \\
& - \i \frac{a_1}{\beta} B_{a0}^2 - \i \frac{a_2}{\beta} B_{ai}^2 - \i \frac{a_8}{\beta} C_{a0}^2 - \i \frac{a_9}{\beta} C_{ai}^2  \label{L_diff}
\end{align}
where all the coefficients in \eqref{L_diff} are purely real due to symmetries summarized in section \ref{variable_symmetry}. Moreover, the constraint \eqref{positive_condition} implies
\begin{align}
a_1 \leq 0, \qquad a_2\leq 0, \qquad a_8 \leq 0, \qquad a_9 \leq 0.
\end{align}
Furthermore, imposing left-right symmetry, we are supposed to have
\begin{align}
a_i = a_{i+7}, \qquad {\rm with} ~~ i = 0,1,2,3,4,5,6
\end{align}

Our result \eqref{L_diff} generalizes relevant ones in the literature in a number of ways. In comparison with \cite{Glorioso:2020loc,Bu:2022esd}, we extend the global symmetry from SU(2) to SU(2)$_L$ $\times$ SU(2)$_R$, and include some higher derivative terms. In comparison with \cite{Hongo:2024brb}, our Lagrangian \eqref{L_diff} contains nonlinear terms hidden in the derivatives that were omitted in \cite{Hongo:2024brb}.

We turn to the Lagrangian $\mathcal{L}_{\Sigma}$ for the order parameter. As for $\mathcal{L}_{diff}$ we retain terms up to quadratic order in the order parameter and to second order in spacetime derivatives. Then, the Lagrangian is
\begin{align}
\mathcal{L}_{\Sigma} & =b_0 \Sigma_{r}^{\dag} \Sigma_a + b_0 \Sigma_{a}^{\dag} \Sigma_r + b_1 \Sigma_a \partial_0 \Sigma_{r}^{\dag} + b_1^* \Sigma_a^{\dag} \partial_0 \Sigma_{r} + b_2 \left(\mathcal{D}_{i} \Sigma_{r} \right)^{\dag} \left(\mathcal{D}_{i} \Sigma_a \right) \nonumber \\
& + b_2 \left( \mathcal{D}_{i} \Sigma_{a} \right)^{\dag} \left(\mathcal{D}_{i} \Sigma_r\right) + b_3 \partial_0 \Sigma_a^\dag \partial_0 \Sigma_r + b_3^* \partial_0 \Sigma_a \partial_0 \Sigma_r^\dag - \i \frac{2 {\rm Re}(b_1)}{\beta} \Sigma_{a}^{\dag}\Sigma_a \nonumber \\
& + \frac{{\rm Im}(b_3)}{\beta} \Sigma_a^\dag \partial_0 \Sigma_a \label{L_SIGMA}
\end{align}
where $b_0$, $b_2$ are purely real, and ${\rm Re}(b_1) \leq 0$. Near the transition point, the coefficient $b_0 \sim (T- T_c)$ with $T_c$ the critical temperature. The fact that $b_1$ and $b_3$ could be complex will be confirmed by the holographic study of section \ref{holo_study}.

To first order in spacetime derivatives, the cubic Lagrangian $\mathcal{L}_{3}$ is\footnote{Indeed, by the KMS symmetry, the $\varpi_{1}$-term shall be accompanied with structures $B_{r0}B_{ri}\partial_iB_{a0}$ and $B_{a0}B_{ri}\partial_{i}B_{r0}$. However, by the chemical shift symmetry, the latter two terms could be written into second order derivatives and thus have been ignored in \eqref{L_BCSIGMA_3}.}
\begin{align}
\mathcal{L}_{3}& = c_{0} \Sigma_r \Sigma_{r}^{\dag} B_{a0} + c_{1} \Sigma_{r} \Sigma_a^{\dag} B_{r0} + c_{1}^* \Sigma_{a} \Sigma_r^{\dag} B_{r0} + \text{i} c_{2} \left(\mathcal{D}_i \Sigma_r \right) \Sigma_{r}^{\dag} B_{ai}  \nonumber \\
& - \text{i} c_{2}^* \Sigma_{r} \left(\mathcal{D}_i\Sigma_r\right)^{\dag} B_{ai} + d_{0} \Sigma_{r}^{\dag} \Sigma_r C_{a0} + d_{1} \Sigma_{a}^{\dag} \Sigma_r C_{r0} + d_{1}^* \Sigma_{r}^{\dag} \Sigma_a C_{r0} \nonumber \\
& + \text{i} d_{2} \Sigma_r^{\dag} \left(\mathcal{D}_i\Sigma_{r}\right) C_{ai}  - \text{i} d_{2}^* \left(\mathcal{D}_i \Sigma_{r}\right)^{\dag} \Sigma_{r} C_{ai} + \varpi_{1} B_{r0} B_{ai} \left( \mathcal{D}_{Li} B_{r0} \right) \nonumber \\
& + \varpi_{2} B_{r0} B_{ai} \partial_0 B_{ri} -\frac{\i\varpi_{2}}{\beta}B_{r0}B_{ai}B_{ai}+ \varpi_{3} C_{r0} C_{ai} \left(\mathcal{D}_{Ri} C_{r0} \right) \nonumber \\
&+ \varpi_{4} C_{r0} C_{ai} \partial_0 C_{ri} - \frac{\i\varpi_{4}}{\beta} C_{r0}C_{ai}C_{ai} + \varpi_{5}B_{a0}B_{r0}B_{r0} + \varpi_{6}C_{a0}C_{r0}C_{r0}   \label{L_BCSIGMA_3}
\end{align}
Here, by $Z_2$ reflection symmetry \eqref{Z2SYM}, the coefficients $c_{0},c_{3},d_{0},d_{3}$ are purely real. Imposing the dynamical KMS symmetry \eqref{KMSsymmetry1}, we have constraints
\begin{equation}
c_{0}=c_{1}=c_{1}^*, \qquad c_{2}=c_{2}^*, \qquad  d_{0}=d_{1}=d_{1}^*, \qquad d_{2}=d_{2}^*
\end{equation}
Interestingly, the Onsager relations among $rrra$-terms \cite{Bu:2024oyz} give an additional constraint:
\begin{equation}
c_2=-d_2=b_2
\end{equation}

Finally, we consider the quartic Lagrangian $\mathcal{L}_{4}$. To zeroth order in spacetime derivatives, the result is
\begin{align}
\mathcal{L}_4 & = \chi_1 \Sigma_{a}^{\dag} \Sigma_r \Sigma_{r}^{\dag} \Sigma_r + \chi_{1}^* \Sigma_{r}^{\dag} \Sigma_a \Sigma_{r}^{\dag} \Sigma_r + c_{3} \Sigma_r \Sigma_{r}^{\dag} B_{r0} B_{a0} + d_{3} \Sigma_{r}^{\dag} \Sigma_r C_{r0} C_{a0} \nonumber \\
& + c_{4} \Sigma_r \Sigma_{a}^{\dag} B_{r0} B_{r0} + d_{4} \Sigma_{a}^{\dag} \Sigma_r C_{r0} C_{r0} + c_{4}^* \Sigma_a \Sigma_{r}^{\dag} B_{r0} B_{r0} + d_{4}^* \Sigma_{r}^{\dag} \Sigma_a C_{r0}C_{r0} \nonumber \\
& + \chi_{2} \Sigma_{r}^{\dag} B_{r0} \Sigma_r C_{a0} + \chi_{3} \Sigma_{r}^{\dag} B_{a0} \Sigma_r C_{r0} + \chi_{4} \Sigma_{r}^{\dag} B_{r0} \Sigma_a C_{r0} + \chi_{4}^{*} \Sigma_{a}^{\dag} B_{r0} \Sigma_r C_{r0} \label{L_BCSIGMA_4}
\end{align}
where the coefficients satisfy
\begin{equation}
\chi_1=\chi_{1}^*,\quad  c_{3}=2c_{4}=2c_{4}^*, \quad d_{3}=2d_{4}=2d_{4}^*, \quad \chi _{4}=\chi _{4}^*.
\end{equation}

In contrast to the EFT for a critical U(1) superfluid \cite{Donos:2023ibv,Bu:2024oyz}, the effective action constructed above contains more fruitful physics thanks to non-Abelian feature for each building blocks. This feature has been recently explored in \cite{Hongo:2024brb} by allowing for weakly explicit breaking of the chiral symmetry.

\subsection{Stochastic equations: non-Abelian model F}

In this section, we derive stochastic equations from the EFT action presented in \eqref{L_diff}, \eqref{L_SIGMA}, \eqref{L_BCSIGMA_3} and \eqref{L_BCSIGMA_4}.

The expectation values of chiral currents are simply obtained by varying $S_{eff}$ with respect to the external sources $\mathcal A_{a\mu}$ and $\mathcal V_{a\mu}$:
\begin{align}
J_L^\mu \equiv \frac{\delta S_{eff}}{\delta \mathcal A_{a\mu}}, \qquad J_R^\mu \equiv \frac{\delta S_{eff}}{\delta \mathcal V_{a\mu}}
\end{align}
The equations of motion for $\varphi_r$ and $\phi_r$ are indeed the conservation laws of chiral currents:
\begin{align}
\frac{\delta S_{eff}}{\delta \varphi_a} = 0 \Rightarrow \partial_\mu J_L^\mu =0, \qquad \frac{\delta S_{eff}}{\delta \phi_a} = 0 \Rightarrow \partial_\mu J_R^\mu =0 \label{eom_J_LR}
\end{align}
Restricted to Gaussian noises, it is equivalent to trade $B_{a\mu}$ and $C_{a\mu}$ in the equations of motion \eqref{eom_J_LR} for noise variables $\xi_L$ and $\xi_R$ \cite{Crossley:2015evo}. Resultantly, \eqref{eom_J_LR} can be rewritten into a stochastic form
\begin{align}
\partial_\mu J^\mu_{L,\, hydro} = \xi_L, \qquad \qquad \partial_\mu J^\mu_{R,\, hydro} = \xi_R, \label{stochastic_J_LR}
\end{align}
where the noises $\xi_L$ and $\xi_R$ obey Gaussian distributions.

The hydrodynamic currents $J^\mu_{L,\, hydro}$ and $J^\mu_{R,\, hydro}$ can be easily read off from the EFT action
\begin{align}
\rho_L & \equiv J_{L, hydro}^{0} \nonumber \\
&= a_0\mu_L + a_1 \partial_0 \mu_L + \left( a_3 + a_5 \right) \tilde{\mathcal{D}}_{Li} \left(\tilde{\mathcal{D}}_{Li} \mu_L - E_{L;i} \right) + a_3 \tilde{\mathcal{D}}_{Li} E_{L;i} \nonumber \\
&\quad +\varpi_5\mu_L^2+ c_1 \mathcal O_r \mathcal O_r^{\dag} + c_3 \mathcal O_r \mathcal O_r^{\dag} \mu_L + \chi_3 \mathcal O_r \mu_R \mathcal O_r^{\dag} \nonumber \\
\rho_R & \equiv J^0_{R,\, hydro} \nonumber \\
& = a_{7}\mu_R + a_8 \partial_0 \mu_R + \left( a_{10} + a_{12} \right) \tilde{\mathcal{D}}_{Ri} \left(\tilde{\mathcal{D}}_{Ri} \mu_R - E_{R;i} \right) + a_{10} \tilde{\mathcal{D}}_{Ri} E_{R;i} \nonumber \\
& \quad +\varpi_6\mu_R^2+ d_{1} \mathcal O_r^{\dag} \mathcal O_r + d_3 \mathcal O_r^{\dag} \mathcal O_r \mu_R + \chi_2\mathcal O_r^{\dag}\mu_L \mathcal O_r \nonumber \\
J^i_{L,\, hydro} & = -a_2 \left( \tilde{\mathcal{D}}_{Li} \mu_L - E_{L;i} \right) + \left(\varpi_1 + \varpi_2 \right) \mu_L \left( \tilde{\mathcal{D}}_{Li} \mu_L - E_{L;i} \right) + a_4 \tilde{\mathcal{D}}_{Li} \partial_0 \mu_L \nonumber \\
& \quad + a_6 \tilde{\mathcal{D}}_{Lj} (\tilde{\mathcal F}_L)_{rij}+ \text{i} c_2 \left[\left( \tilde{\mathcal{D}}_i \mathcal O_r \right) \mathcal O_r^{\dag} - \mathcal O_r \left( \tilde{\mathcal{D}}_i \mathcal O_r \right)^{\dag} \right] + \varpi_1 \mu_L E_{L;i} \nonumber \\
J^i_{R,\, hydro} & = -a_9 \left( \tilde{\mathcal{D}}_{Ri} \mu_R - E_{R;i} \right) + \left(\varpi_3 + \varpi_4 \right) \mu_R \left( \tilde{\mathcal{D}}_{Ri} \mu_R - E_{R;i} \right) + a_{11} \tilde{\mathcal{D}}_{Ri} \partial_0 \mu_R \nonumber \\
& \quad + a_{13} \tilde{\mathcal{D}}_{Rj} (\tilde{\mathcal F}_R)_{rij} + \text{i} d_2 \left[\left( \tilde{\mathcal{D}}_i \mathcal O_r \right) \mathcal O_r^{\dag} - \mathcal O_r \left( \tilde{\mathcal{D}}_i \mathcal O_r \right)^{\dag} \right] + \varpi_4 \mu_R E_{R;i} \label{J_LR_hydro}
\end{align}
Here, the chemical potentials $\mu_L, \mu_R$ and the order parameter $\mathcal O_r$ are defined as\footnote{Indeed, the last equation follows from the definition of \eqref{Sigma}.} \cite{Glorioso:2020loc}
\begin{align}
\mu_L \equiv \mathcal U^\dag (\varphi_r) B_{r0} \mathcal U(\varphi_r), \qquad \mu_R \equiv \mathcal U^\dag (\phi_r) C_{r0} \mathcal U(\phi_r), \qquad \mathcal O_r \equiv \mathcal U^\dag (\varphi_r) \Sigma_r \mathcal U(\phi_r)
\end{align}
The $E_{Li}$, $(\tilde{\mathcal F}_L)_{rij}$, $E_{Ri}$ and $ (\tilde{\mathcal F}_R)_{rij}$ are electromagnetic fields associated with the background non-Abelian gauge fields $\mathcal A_{r\mu}$ and $\mathcal V_{r\mu}$. The derivative operators in \eqref{J_LR_hydro} are obtained from \eqref{covariant_derivative_spatial} by replacing $B_{ri}$ and $C_{ri}$ by the background fields $\mathcal A_{ri}$ and $\mathcal V_{ri}$:
\begin{align}
\tilde{\mathcal{D}}_{Li} \equiv \partial_{i}-\text{i}[\mathcal A_{ri},\cdot], \qquad \tilde{\mathcal{D}}_{Ri} \equiv \partial_{i}-\text{i}[\mathcal V_{ri},\cdot], \qquad
\tilde{\mathcal{D}}_{i} \equiv  \partial_i-\text{i} \mathcal A_{ri}\cdot + \text{i}\cdot \mathcal V_{ri} \label{covariant_derivative_external}
\end{align}
Interestingly, \eqref{J_LR_hydro} generalizes the U(1) charge diffusion to non-Abelian situation, with contribution from a charged order parameter included. This demonstrates the two-fluid picture for superfluid.

In the same spirit, treating $\Sigma_a$ as a noise variable $\zeta$, we obtain a stochastic equation for the order parameter:
\begin{align}
\frac{\delta S_{eff}}{\delta \Sigma_a^{\dag}} =0 \Rightarrow \frac{J_{\mathcal O}}{b_1^*} = \zeta, \label{stochastic_O}
\end{align}
where
\begin{align}
J_{\mathcal O} = & b_0 \mathcal O_r + b_1^* \left( \partial_0 \mathcal O_r + \i\, \mu_L \mathcal O_r -\i\, \mathcal O_r \mu_R \right)- b_2 \tilde{\mathcal{D}}_{i}^{\dag}\left( \tilde{\mathcal{D}}_i \mathcal O_r \right) + c_0 \mu_L \mathcal O_r + d_0 \mathcal O_r \mu_R \nonumber \\
& + \chi_1 \mathcal O_r \mathcal O_r^{\dag} \mathcal O_r + c_4 \mu_L \mu_L \mathcal O_r + d_4 \mathcal O_r \mu_R \mu_R + \chi _4 \mu_L \mathcal O_r \mu_R \label{J_O}
\end{align}

In deriving \eqref{J_LR_hydro} and \eqref{J_O}, we have ignored second order time-derivative terms in the EFT action. This is valid for rewriting the equations of motion \eqref{stochastic_J_LR} and \eqref{stochastic_O} into ``non-Abelian'' model F in the Hohenberg-Halperin classification, 
which involves only {\it first-order time-derivatives}. Physically, this is motivated by the scaling $\partial_0 \sim \partial_i^2$ in the symmetric phase \cite{RevModPhys.49.435}. Therefore, by neglecting second order time-derivatives in the EFT action, the stochastic equations will not cover those effects arising from quartic spatial derivatives.

The reason of considering the model F is mainly inspired by the formally significant similarity between the QCD chiral phase transition and the U(1) superfluid phase transition, with the latter belonging to the model F. Essentially, both are related to spontaneous breaking of continuous global symmetries. So, in this
sense, we view the two-flavor QCD in the chiral limit as a non-Abelian superfluid (belonging to the ``non-Abelian'' model F). Indeed, our stochastic equations can be nicely recast into that of the O(4) model G by the following treatments: combing $\rho_L$ and $\rho_R$ into an O(4) traceless symmetric tensor (in the flavor space); properly ignoring higher-order terms.

The equations \eqref{stochastic_J_LR} and \eqref{stochastic_O} are stochastic equations for the chemical potentials $\mu_{L,R}$ and the chiral condensate $\mathcal O_r$. We advance by trading $\mu_{L,R}$ for $\rho_{L,R}$, which makes it more convenient to compare our results with \cite{RevModPhys.49.435}. Inverting the first two equations in \eqref{J_LR_hydro}, we are supposed to get functional relations
\begin{align}
\mu_L = \mu_L[\rho_L, \rho_R, \mathcal O_r], \qquad \mu_R = \mu_R[\rho_L, \rho_R, \mathcal O_r]
\end{align}
which help to rewrite equations of motion \eqref{stochastic_J_LR} and \eqref{stochastic_O} into stochastic equations for the charge densities and chiral condensate. 
In order to ease the matching between the equations of motion derived from our EFT action and the stochastic equations of \cite{RevModPhys.49.435} (in which external sources for the chiral currents were not considered), we switch off the external fields $\mathcal A_{r\mu}$ and $\mathcal V_{r\mu}$. By this simplification, the equations of motion will not contain source terms. In principle, they may be recovered by replacing the spacetime derivatives by gauge covariant derivatives.
%
%
Eventually, the stochastic equations are
\begin{align}
\partial_0 \rho_L & = \frac{a_2}{a_0} \nabla^{2} \rho_L - \frac{a_2 c_1} {a_0} \nabla^{2} \left( \mathcal O_r \mathcal O_r^{\dag} \right) - \text{i} c_2 \left[ \left(\nabla^2 \mathcal O_r \right)\mathcal O_r^{\dag} - \left(\nabla^2 \mathcal O_r^{\dag}\right) \mathcal O_r \right]\nonumber \\
& - \left(\frac{\varpi_1 + \varpi_{2}} {2a_0^2}+\frac{a_2\varpi_5}{a_0^3}\right) \nabla^{2} \rho_L^2 + \xi_L \nonumber \\
\partial_0 \rho_R & = \frac{a_8}{a_{7}} \nabla^{2} \rho_R  - \frac{a_8 d_1}{2a_7} \nabla^{2} \left(\mathcal O_r \mathcal O_r^{\dag} \right) - \text{i} d_2 \left[ \left(\nabla^2 \mathcal O_r \right)\mathcal O_r^{\dag} - \left(\nabla^2 \mathcal O_r^{\dag}\right) \mathcal O_r \right]  \nonumber \\
&- \left(\frac{\varpi_3 + \varpi_{4}} {2a_7^2}+\frac{a_8\varpi_6}{a_7^3}\right) \nabla^{2} \rho_R^2 + \xi_R \nonumber \\
\partial_0 \mathcal O_r & = \frac{b_0}{b_1^*} \mathcal O_r - \frac{b_2}{b_1^*} \nabla^2 \mathcal O_r + \frac{c_0}{b_1^*a_0} \rho_L \mathcal O_r + \frac{d_0}{b_1^*a_7} \mathcal O_r \rho_R + \frac{1}{b_1^*} \left( \chi_1 + \frac{c_0^2}{a_0} + \frac{d_0^2}{a_7} \right) \mathcal O_r \mathcal O_r^{\dag}\mathcal O_r \nonumber \\
& - \i \left( \frac{1}{a_0} \rho_L \mathcal O_r - \frac{1}{a_7} \mathcal O_r \rho_R \right) + \i \left( \frac{c_1}{a_0} - \frac{d_1}{a_7} \right) \mathcal O_r \mathcal O_r^\dag \mathcal O_r + \left(\frac{c_4}{b_1^*a_0^2}-\frac{c_0\varpi_5}{b_1^*a_0^3}\right) \mathcal O_r \rho_L \rho_L  \nonumber \\
& + \left(\frac{d_4}{b_1^*a_7^2}-\frac{d_0\varpi_6}{b_1^*a_7^3}\right) \mathcal O_r \rho_R \rho_R+ \frac{\chi_4}{b_1^*a_0a_7} \mathcal O_r \rho_L \rho_R + \zeta \label{stochastic_rho_O}
\end{align}

Presumably, the effective theory we constructed corresponds to a non-Abelian superfluid near the critical temperature. It is then of interest to compare the set of equations \eqref{stochastic_rho_O} with that of the model F under the Hohenberg-Halperin classification \cite{RevModPhys.49.435}, with the latter an effective description for U(1) superfluid near the critical point. We find that, with the terms $\nabla^2 \rho_{L,R}^2$, $\mathcal O_r \rho_L \rho_L$, $\mathcal O_r \rho_R \rho_R$ and $\mathcal O_r \rho_L \rho_R$ ignored, our results \eqref{stochastic_rho_O} can be viewed as non-Abelian version of the stochastic equations of the model F. Intriguingly, the $\nabla^2 \rho_{L,R}^2$ terms in the evolution equation of $\rho_{L,R}$ resemble Kardar-Parisi-Zhang (KPZ) term \cite{Kardar:1986xt}, which has been unveiled from the EFT perspective in \cite{Crossley:2015evo}. The terms of the form $\mathcal O_r \rho \rho$ in the evolution equation of $\mathcal O_r$ represent higher order terms given that $\rho \sim \mathcal O^2$ near the phase transition. However, it is important to stress that the EFT approach provides a systematic way of generalizing the widely used stochastic models.

Before concluding this section, we briefly discuss spontaneous $\chi$SB based on the EFT presented above. Recall that below the critical temperature $T_c$, the coefficient $b_0 $ becomes negative. Then, from \eqref{stochastic_rho_O}, we immediately conclude that a stable homogeneous configuration in the low temperature phase can be taken as
\begin{align}
\mathcal O_r = \bar{\mathcal O} \neq 0, \qquad \rho_L =0, \qquad \rho_R =0 \label{condensed_state}
\end{align}
where $\bar{\mathcal O}$ is a constant background for the chiral condensate operator, characterizing spontaneous $\chi$SB. Now, we consider perturbations on top of the state \eqref{condensed_state}
\begin{align}
\mathcal O_r = \left(\bar{\mathcal O} + \delta \mathcal O \right) e^{\i \theta} \approx \bar{\mathcal O} + \delta \mathcal O + \i \bar{\mathcal O} \theta, \qquad \rho_L = 0 + \delta \rho_L, \qquad \rho_R = 0 + \delta \rho_R \label{perturb_lowT}
\end{align}
Plugging \eqref{perturb_lowT} into \eqref{stochastic_rho_O} and keeping linear terms in the perturbations, we will find \cite{Donos:2022qao,RevModPhys.49.435} propagating modes (Goldstone modes) of the form $\theta + \delta \rho_L - \delta \rho_R$. These are non-Abelian generalization of the U(1) superfluid sound mode and correspond to the pions associated with spontaneous $\chi$SB. Beyond linear level, we are supposed to have an interacting theory for density variations $\delta \rho_{L,R}$ and chiral condensate variation $\delta \mathcal O, ~ \theta$. In fact, it will be interesting to carry out such an analysis based on the EFT action, yielding a generalized chiral perturbation theory valid for finite temperature \cite{Grossi:2021gqi}. We leave this interesting exploration as future work.

\section{Holographic EFT for a modified AdS/QCD} \label{holo_study}

In this section, we confirm the EFT action constructed in section \ref{EFT_constru} through a holographic study for a modified AdS/QCD model.

\subsection{Holographic setup} \label{holographic_setup}


Holographically mimicking the QCD matter involves an action in a five dimensional spacetime
\begin{align}
S_{\rm bulk} = S_{\rm gra} + S_0 \label{S_bulk}
\end{align}
where $S_{\rm gra}$ and $S_0$ are dual to the gluonic sector and flavor sector of QCD, respectively. A widely used model for $S_{\rm gra}$ consists of Einstein-dilaton with a specifically-chosen potential for the dilaton field \cite{Gubser:2008ny,Gubser:2008yx,Gursoy:2008bu,Gursoy:2008za}. The effects of finite baryon density and background magnetic field could be accounted for by adding a U(1) gauge field \cite{DeWolfe:2010he,Finazzo:2016mhm,Knaute:2017opk,Critelli:2017oub} to the setup of \cite{Gubser:2008ny,Gubser:2008yx,Gursoy:2008bu,Gursoy:2008za}. For the flavor sector, we use
the modified AdS/QCD model \cite{Chelabi:2015gpc,Chelabi:2015cwn,Chen:2018msc,Cao:2022csq}
\begin{equation}
S_0=\int{d^5}x\sqrt{-g}\, \text{Tr}\Bigl\{ -|DX|^2 - \left( m_0^2 - \frac{\mu^2}{r^2} \right) |X|^2 -a |X|^4 - \frac{1}{4} \left( F_{L}^{2}+F_{R}^{2} \right) \Bigr\}    \label{5Dbulkaction}
\end{equation}
Here, the scalar field $X$ is dual to the chiral condensate $\mathcal O$. Thus, $X$ is in the fundamental representation of bulk $SU(2)_L \times SU(2)_R$ gauge symmetry, and $D_M X = \nabla_M X - \i A_{LM} X + \i X A_{R M}$. The $SU(2)_L$ gauge potential $A_{LM} = A_{LM}^\a t^\a$ is dual to the left-handed current $J_L^\mu$ and $A_{RM}$ dual to $J_R^\mu$. $F_L$ denotes field strength of $SU(2)_L$ Yang-Mills field $A_L$ with $F_{L M N} = \nabla_M A_{L N} - \nabla_N A_{L M} - \i \left[ A_{LM},A_{LN} \right]$ and similarly for $F_R$.

The above setup can be straightforwardly extended to study the (2+1)-flavor QCD. The bulk gauge symmetry $SU(2)_L \times SU(2)_R$ will be enlarged to $U(3)_L \times U(3)_R$. The vectorial subgroup U(1) represents the baryon number symmetry. The finite quark masses are realized via a matrix-valued source (i.e., the non-normalizable mode) in the bulk scalar field. With the mass difference between the $u$- and $d$-quarks neglected, the vectorial subgroup $SU_V(3)$ is explicitly broken to $SU(2)_I \times U(1)_S$, corresponding to the isospin and strangeness, respectively.

The original AdS/QCD model \cite{Erlich:2005qh} corresponds to setting $\mu=0$ in \eqref{5Dbulkaction}, which does not incorporate spontaneous $\chi$SB. This shortcoming was resolved by introducing the $\mu$-term in the scalar mass \cite{Chelabi:2015gpc,Chelabi:2015cwn,Chen:2018msc,Cao:2022csq}. 
As discussed in \cite{Cao:2022csq}, this extra term might originate from the coupling of the scalar field to an extra field, say $\phi |X|^2$, where $\phi$ might be a dilaton field having a non-trivial background $\phi \sim 1/r^2$. Treating in this way, we would inevitably have to consider the dynamics of the $\phi$ (which also couples to the bulk metric field). This is far beyond the scope of present study. Thus, we take an effective approach and view the $r$-dependent mass term as a phenomenological input. We would like to point out that such a treatment does not violate basic rules like translational symmetry, since the background spacetime explicitly breaks the translational invariance along the $r$-direction.

We take $m_0^2 = - 3/L^2$ with $L$ the AdS radius so that the dual operator $\mathcal O$ has a scaling dimension three\footnote{ Since the extra term $\mu^2/r^2$ represents a higher order term near the AdS boundary, it does not affect the scaling dimension of the dual operator $\mathcal O$.} as required for real-world QCD. In this work we will focus on the dynamics near the phase transition. Then, the phenomenological parameter $\mu$ in \eqref{5Dbulkaction} can be written as
\begin{align}
\mu = \mu_c + \delta \mu \label{mu_expansion}
\end{align}
Here, $\mu_c$ is a critical value of $\mu$ at which the order parameter $\mathcal O$ vanishes when its external source is zero. Numerically, this condition gives $\mu_c = 2.40 r_h^2$. The $\delta \mu$ stands for a tiny deviation from the critical value $\mu_c$. Throughout this work we will set the AdS radius $L$ to unity.


Recall that the EFT presented in section \ref{EFT_constru} focuses on the dynamics of flavor sector around a thermal state with zero baryon density. In particular, the variations of energy and momentum densities are turned off. This corresponds to the probe limit on the bulk side which we explain here. We are supposed to use the $S_{\rm gra}$ to obtain a static black brane solution, which is dual to the QCD thermal state. Then, we will study the dynamics of \eqref{5Dbulkaction} in this static black brane background. In particular, in parallel with the EFT construction, we do not turn on perturbations for the bulk metric and dilaton fields.

However, the static black brane solution obtained from $S_{\rm gra}$ is known numerically only \cite{Gubser:2008ny,Gubser:2008yx,Gursoy:2008bu,Gursoy:2008za}, which will make the derivation of dual EFT action rather challenging. Technically, this is owing to the requirement of systematically including both the ingoing mode (dual to the dissipation) and outgoing mode (dual to the thermal fluctuation) for all the matter fields in \eqref{5Dbulkaction}. Therefore, as a qualitative study, we follow \cite{Cao:2022csq} and take the Schwarzschild-AdS$_5$ black brane as a substitute for the numerical one obtained from $S_{\rm gra}$. Notice that the Schwarzschild-AdS$_5$ geometry is dual to $\mathcal N =4$ SU($N_c$) super-Yang-Mills theory at finite temperature, which is very different from thermal QCD at the microscopic level. Nevertheless, from the perspective of Wilsonian renormalization group (RG), we know that systems that show remarkable differences at microscopic scale may flow to the same infrared (IR) fixed point (the critical point), and thus belong to the same universality class. Given that the dual model \eqref{5Dbulkaction} correctly captures the flavor symmetry of two-flavor QCD, we believe that the dual EFT action to be derived will take the same form as the one presented in section \ref{EFT_constru}. Our holographic study will clearly demonstrate this expectation. We further point out that the holographic values for various coefficients in the dual EFT are quite sensitive to the thermal state under consideration, and might be quite different from those of real QCD. In other words, if we had used the numerical black brane geometry obtained from $S_{\rm gra}$, we expect to obtain the same form of dual EFT action but with different values for various coefficients reflecting thermal state of QCD. We leave the ambitious study of deriving dual EFT based on the whole bulk system \eqref{S_bulk} as a future project.

\begin{figure}[htbp]
\centering
\includegraphics[width=1\textwidth]{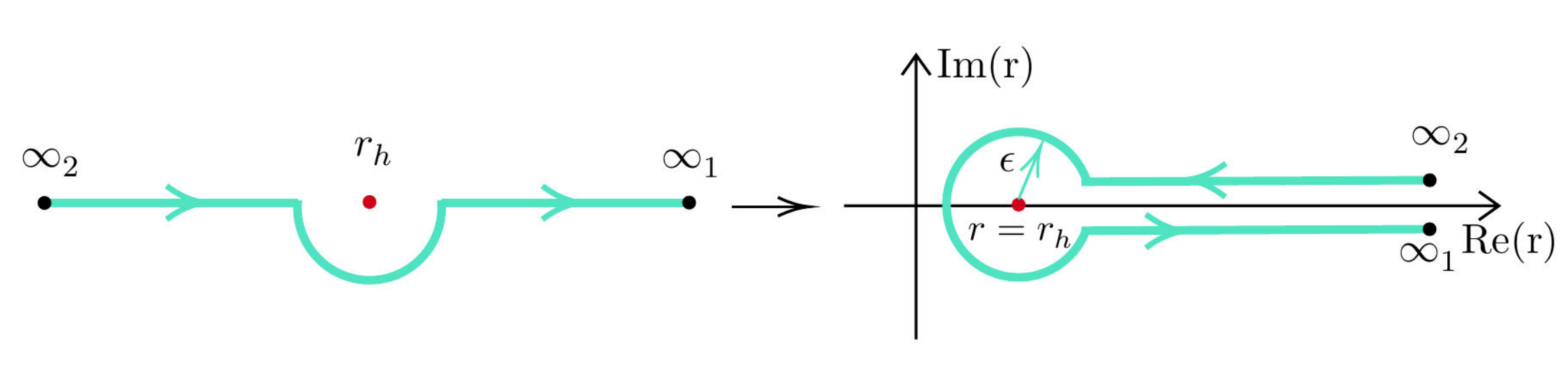}
\caption{Left: complexified double AdS (analytically continued near the horizon) \cite{Crossley:2015tka}; Right: the holographic SK contour \cite{Glorioso:2018mmw}. The two horizontal legs overlap with the real axis.}
\label{Holographic_path}
\end{figure}

Eventually, in the ingoing Eddington-Finkelstein (EF) coordinate system $x^M=(r,v,x^i)$, the metric of background geometry is given by
\begin{equation}
ds^2=g_{MN}dx^{M}dx^{N}=2dvdr-r^2f\left( r \right) dv^2+r^2\delta _{ij}dx^idx^j, \qquad i,j=1,2,3  \label{Schw-AdS5}
\end{equation}
where $f\left( r \right) =1-{r^4_h}/{r^4}$ with $r_h$ the horizon radius. The Schwarzschild-AdS$_5$ \eqref{Schw-AdS5} has a Hawking temperature $T={r_h}/\pi$, which is identified as the temperature for the boundary theory. In order to derive the effective action for the boundary theory, we apply the holographic SK technique \cite{Glorioso:2018mmw} in which the radial coordinate varies along a contour of Figure \ref{Holographic_path}. 
The reason of using the ingoing EF coordinate system is to remove coordinate singularity, which is crucial in forming the radial contour of Figure \ref{Holographic_path}.

The bulk equations of motion derived from \eqref{5Dbulkaction} are
\begin{align}
EL^M=0, \qquad ER^M=0, \qquad EX =0 \label{bulk_eom}
\end{align}
where
\begin{align}
EL^M &= \nabla_N \left( F_L \right)^{MN} + \i \left[ A_{LN},\left( F_L \right)^{NM} \right] + (\mathcal J_L)^M \nonumber \\
ER^M &= \nabla_N \left( F_R \right)^{MN} + \i \left[ A_{RN},\left( F_R \right)^{NM} \right] + (\mathcal J_R)^M \nonumber \\
EX &= D^M (D_M X) - \left( m_0^2 - \frac{\mu^2}{r^2} \right) X - 2 a \left( X^{\dag} X \right) X
\end{align}
The bulk currents are
\begin{align}
( {\mathcal J}_L )^M & = -\i X \left( D^M X \right)^{\dag} + \i \left(D_M X \right) X^{\dag}, \nonumber \\
( {\mathcal J}_R )^M & = -\i X^\dag \left( D^M X \right) + \i \left(D_M X \right)^{\dag} X
\end{align}

The bulk gauge symmetry allows us to take the following radial gauge condition \cite{Bu:2020jfo}
\begin{align}
A_{Lr}=-\frac{A_{L0}}{r^2f(r)}, \qquad \qquad A_{Rr}=-\frac{A_{R0}}{r^2f(r)}. \label{gaugefixing}
\end{align}
Then, near the AdS boundary, the bulk fields behave as
\begin{align}
A_{L\mu}\left( r \to \infty_s,\ x^{\alpha} \right) &= B_{s\mu}(x^\alpha) + \cdots + \frac{\mathfrak J_{s\mu}(x^\alpha)}{r^2} + \cdots, \nonumber \\
A_{R\mu}\left( r \to \infty_s,\ x^{\alpha} \right) &= C_{s\mu}(x^\alpha) + \cdots + \frac{\mathfrak L_{s\mu}(x^\alpha)}{r^2} + \cdots, \nonumber \\
X\left( r \to \infty_s,\ x^{\alpha} \right) &= \frac{m_s(x^\alpha)}{r} + \cdots + \frac{\Sigma_s(x^\alpha)}{r^3} + \cdots,  \label{AdSboundarycondition}
\end{align}
where $B_{s\mu}$, $C_{s\mu}$ and $\Sigma_{s}$ are exactly the dynamical variables introduced in section \ref{EFT_constru} for writing down the EFT action. Therefore, when solving bulk equations \eqref{bulk_eom}, we will impose boundary conditions as follows: take $B_{s\mu}$, $C_{s\mu}$ and $\Sigma_{s}$ as boundary data and fix them. So, once bulk equations are solved, the rest modes in \eqref{AdSboundarycondition} will become functionals of the boundary data.

It turns out that in order to fully determine the bulk gauge fields, we have to impose extra boundary conditions at the horizon \cite{Glorioso:2018mmw}
\begin{align}
A_{L0}(r=r_h -\epsilon, x^\alpha) = 0, \qquad A_{R0}(r=r_h -\epsilon, x^\alpha) = 0  \label{horizon_condition}
\end{align}
which further break the residual gauge invariance of bulk theory after taking the radial gauge condition \eqref{gaugefixing}. Physically, the horizon conditions \eqref{horizon_condition} correspond to chemical shift symmetry for the boundary theory.


With the near boundary asymptotic behaviors \eqref{AdSboundarycondition}, it is straightforward to show that the bulk action \eqref{5Dbulkaction} contains divergences near the AdS boundary. These divergences can be systematically removed by the standard procedure of holographic renormalization \cite{Skenderis:2002wp}: regulating the bulk theory by placing a cutoff near the AdS boundary, and adding suitable counter-term action to cancel the divergences. Skipping the details, we present the suitable counter-term action \cite{Round:2010kj,Bu:2022esd,Marolf:2006nd}
\begin{align}
S_{ct} = S_{ct}^A + S_{ct}^X
\end{align}
where
\begin{align}
& S_{ct}^A = \frac{1}{4}\log r\, \int d^4x \sqrt{-\gamma} {\rm Tr} \left( F_{L\mu\nu} F_L^{\mu\nu} + F_{R\mu\nu} F_R^{\mu\nu} \right), \nonumber \\
& S_{ct}^X=\int{d^4x}\sqrt{-\gamma}\, {\rm Tr} \left[ n_M X^{\dag} D^M X + n_M X (D^M X)^{\dag}+X^{\dag}X  - (\overline{\nabla }_{\mu}X)^{\dag} (\overline{\nabla }^{\mu}X ) \log r \right], \label{Sct_A_X}
\end{align}
where $r\to \infty$ is assumed. The counter-term action \eqref{Sct_A_X} is written down in the minimal subtraction scheme. Here, $\gamma$ is the determinant of induced metric $\gamma_{\mu\nu}$ on the boundary $r=\infty$, $n_M$ is the normal vector of the boundary hypersurface with $r=\infty$, and $\overline{\nabla }_\mu$ is the 4D covariant derivative compatible with the induced metric $\gamma_{\mu\nu}$.

However, it turns out that the variational problem based on $S_0 + S_{ct}$ is not well-defined. That is to say, provided the boundary condition specified below \eqref{AdSboundarycondition}, the variation $\delta (S_0 + S_{ct} )$ does not vanish on shell (i.e., with bulk EOMs imposed). In order to cure this issue, we need to add a boundary term for the scalar sector \cite{Bu:2021clf,Bu:2024oyz}
\begin{equation}
S_{bdy}=\int{d^4x}\mathrm{Tr}\left(\frac{1}{2}m^{\dag}\Box m -m^{\dag}\partial_0^2 m\right), \label{S_bdy}
\end{equation}
where $\Box = \eta^{\mu\nu}\partial_\mu \partial_\nu$ with $\eta^{\mu\nu}$ the 4D Minkowski metric. Eventually, the on-shell variation of the total bulk action reads
\begin{align}
\delta (S_0 + S_{ct} + S_{bdy}) = \int d^4x {\rm Tr}\left( \mathfrak J_s^\mu \delta B_{s\mu} + \mathfrak L_s^\mu \delta C_{s\mu} -2 m_s^\dag \delta \Sigma_s -2 m_s \delta \Sigma_s^\dag \right)
\end{align}
which apparently vanishes given the boundary conditions specified below \eqref{AdSboundarycondition}. This demonstrates that the bulk variational problem based on $S_0 + S_{ct} + S_{bdy}$ is well-defined.

In the saddle point approximation, the derivation of boundary effective action boils down to solving classical equations of motion for the bulk theory \eqref{bulk_eom}. However, to ensure the dynamical variables encoded in the boundary data off-shell, we will adopt a partially on-shell approach to solve the bulk dynamics, as demonstrated from bulk partition function \cite{Crossley:2015tka,Bu:2021clf,Bu:2024oyz}. Therefore, under the radial gauge choice \eqref{gaugefixing}, we will solve the dynamical components of bulk equations \cite{Bu:2024oyz,Crossley:2015tka}
\begin{align}
EL^0 - \frac{EL^r}{r^2 f(r)} =0, \qquad ER^0 - \frac{ER^r}{r^2 f(r)} =0, \qquad EL^i =0, \qquad  ER^i =0, \qquad EX=0 \label{dynamical_eom}
\end{align}
while leave aside the constraint equations
\begin{align}
EL^r =0, \qquad ER^r=0
\end{align}
The boundary effective action is identified as
\begin{align}
S_{eff} = S_0|_{p.o.s} + S_{ct} + S_{bdy}
\end{align}
where $S_0|_{p.o.s}$ stands for the partially on-shell bulk action obtained by plugging solutions for \eqref{dynamical_eom} into the bulk action \eqref{5Dbulkaction}.

\subsection{Bulk perturbation theory} \label{perturb_theory}

In this section we set up a perturbative approach for solving the dynamical equations \eqref{dynamical_eom}. Recall that the EFT action presented in section \ref{EFT_constru} is organized by the number of dynamical fields $B_{s\mu}$, $C_{s\mu}$ and $\Sigma_s$ as well as the number of spacetime derivatives of these fields. Accordingly, our strategy of solving \eqref{dynamical_eom} will be through a double expansion.

First, we expand the bulk fields as
\begin{align}
A_{LM} &= \alpha A_{LM}^{(1)} + \alpha^2 A_{LM}^{(2)} + \cdots, \qquad A_{RM} = \alpha A_{RM}^{(1)} + \alpha^2 A_{RM}^{(2)} + \cdots, \nonumber \\
X &= \alpha X^{(1)} + \alpha^2 X^{(2)} + \cdots, \label{alpha-expansion}
\end{align}
where the bookkeeping parameter $\alpha$ assists in counting the number of dynamical variables $B_{s\mu}$, $C_{s\mu}$ and $\Sigma_s$. This can be viewed as linearization over the highly nonlinear system \eqref{dynamical_eom}. Indeed, the leading order solutions $A_{LM}^{(1)}$ and $A_{RM}^{(1)}$ obey free Maxwell equations in the background spacetime \eqref{Schw-AdS5}. The nonlinear solutions like $A_{LM}^{(2)}$ and $A_{RM}^{(2)}$ obey similar equations as those of $A_{LM}^{(1)}$ and $A_{RM}^{(1)}$, with nontrivial sources to be built from lower order solutions. An analogous statement applies to the scalar sector $X$: the leading part $X^{(1)}$ satisfies free Klein-Gordon (KG) equation in Schwarzschild-AdS$_5$, while the nonlinear parts like $X^{(2)}$ obey inhomogeneous KG equation with sources constructed from lower order solutions.

Next, at each order in the expansion \eqref{alpha-expansion}, we do a boundary derivative expansion
\begin{align}
A_{LM}^{(m)} &= A_{LM}^{(m)(0)} + \lambda A_{LM}^{(m)(1)} + \lambda^2 A_{LM}^{(m)(2)} + \cdots, \nonumber \\
A_{RM}^{(m)} &= A_{RM}^{(m)(0)} + \lambda A_{RM}^{(m)(1)} + \lambda^2 A_{RM}^{(m)(2)} + \cdots, \nonumber \\
X^{(m)} &= X^{(m)(0)} + \lambda X^{(m)(1)} + \lambda^2 X^{(m)(2)} + \cdots, \label{lambda-expansion}
\end{align}
where $\lambda$ helps to count the number of boundary derivatives.

Thanks to the double expansion \eqref{alpha-expansion} and \eqref{lambda-expansion}, the dynamical equations \eqref{dynamical_eom} turn into a set of linear ordinary differential equations (ODEs) which we schematically write here
\begin{align}
&\Box_0 A_{L0}^{(m)(n)} = j_{L0}^{(m)(n)}, \qquad \Box_i A_{Li}^{(m)(n)} = j_{Li}^{(m)(n)}, \nonumber \\
&\Box_0 A_{R0}^{(m)(n)} = j_{R0}^{(m)(n)}, \qquad \Box_i A_{Ri}^{(m)(n)} = j_{Ri}^{(m)(n)}, \nonumber \\
&\Box_X X^{(m)(n)} = j_X^{(m)(n)}  \label{dynamical_eom_perturb}
\end{align}
where the differential operators can be read off from \eqref{dynamical_eom} by ignoring the boundary spacetime derivatives
\begin{align}
\Box_0 = \partial_r(r^3\partial_r ), \qquad \Box_i = \partial_r[r^3f(r) \partial_r], \qquad \Box_X = \partial_r [ r^5f(r) \partial_r] - \left(m_0^2 - \frac{\mu^2}{r^2} \right)
\end{align}
The source terms are easily read off by plugging the double expansion \eqref{alpha-expansion} and \eqref{lambda-expansion} into the dynamical equations \eqref{dynamical_eom}.

{\bf Perturbative solutions for the gauge sector}

For the gauge sector, we can recycle our previous results for perturbative solutions \cite{Baggioli:2023tlc,Bu:2024oyz}. For the leading order parts, we have
\begin{align}
& A_{L0}^{(1)(0)}(r) = B_{s 0} \left(1- \frac{r_h^2}{r^2} \right), \qquad r\in[r_h - \epsilon, \infty_s), \nonumber \\
& A_{R0}^{(1)(0)}(r) = C_{s 0} \left(1- \frac{r_h^2}{r^2} \right), \qquad r\in[r_h - \epsilon, \infty_s), \nonumber \\
& A_{Li}^{(1)(0)}(r) = B_{2i} + \frac{B_{ai}}{2\i \pi} \log \frac{r^2 -r_h^2}{r^2 + r_h^2}, \nonumber \\
& A_{Ri}^{(1)(0)}(r) = C_{2i} + \frac{C_{ai}}{2\i \pi} \log \frac{r^2 -r_h^2}{r^2 + r_h^2} \label{A_LR_00}
\end{align}

For the next to leading order parts, we have
\begin{align}
A_{L0}^{(1)(1)}(r) &= \frac{\partial_0 B_{s 0}}{4r_h} \left( 1- \frac{r_h^2}{r^2} \right) \left[ \pi - 2 \arctan\left( \frac{r}{r_h}\right) + \log \frac{r+r_h}{r-r_h} \right], \quad r\in[r_h-\epsilon, \infty_s), \nonumber \\
A_{R0}^{(1)(1)}(r) &= \frac{\partial_0 C_{s 0}}{4r_h} \left( 1- \frac{r_h^2}{r^2} \right) \left[ \pi - 2 \arctan\left( \frac{r}{r_h}\right) + \log \frac{r+r_h}{r-r_h} \right], \quad r\in[r_h-\epsilon, \infty_s), \nonumber \\
A_{Li}^{(1)(1)}(r) &= \frac{\partial_0 B_{2i}}{4r_h} \left[ \pi - 2 \arctan\left( \frac{r}{r_h} \right) + 2\log(r+r_h) - \log (r^2 +r_h^2) \right] \nonumber \\
& - \frac{\partial_0 B_{a i}}{8\pi r_h} \left[ -(2-\i) \pi - 2 \i \arctan\left( \frac{r}{r_h} \right) - \i \log \frac{r-r_h}{r+r_h}  \right] \log \frac{r^2 -r_h^2}{r^2 + r_h^2}, \nonumber \\
A_{Ri}^{(1)(1)}(r) &= \frac{\partial_0 C_{2i}}{4r_h} \left[ \pi - 2 \arctan\left( \frac{r}{r_h} \right) + 2\log(r+r_h) - \log (r^2 +r_h^2) \right] \nonumber \\
& - \frac{\partial_0 C_{a i}}{8\pi r_h} \left[ -(2-\i) \pi - 2 \i \arctan\left( \frac{r}{r_h} \right) - \i \log \frac{r-r_h}{r+r_h}  \right] \log \frac{r^2 -r_h^2}{r^2 + r_h^2}. \label{A_LR_01}
\end{align}

For the higher order solutions, instead of recording lengthy expressions, we write them compactly as radial integrals. For the time-components, we have
\begin{align}
& A_{L0}^{(m)(n)}(r) = \int_{\infty_s}^r \left[ \frac{1}{x^3} \int_{\infty_s}^x j_{L0}^{(m)(n)}(y) dy + \frac{c_s^{(m)(n)}}{x^3} \right] dx, \quad r\in[r_h-\epsilon, \infty_s), \nonumber \\
& A_{R0}^{(m)(n)}(r) = \int_{\infty_s}^r \left[ \frac{1}{x^3} \int_{\infty_s}^x j_{R0}^{(m)(n)}(y) dy + \frac{d_s^{(m)(n)}}{x^3} \right] dx, \quad \,\, r\in[r_h-\epsilon, \infty_s), \label{A_LR00_mn}
\end{align}
where the integration constants $c_s^{(m)(n)}$ and $d_s^{(m)(n)}$ are determined by the horizon conditions \eqref{horizon_condition}.

For the spatial components, we have
\begin{align}
A_{Li}^{(m)(n)}(r) &= \int_{\infty_2}^{\infty_1} G_Y(r,\xi) j_{Li}^{(m)(n)}(\xi) d\xi  \nonumber \\
& = \frac{Y_1(r)}{2\i\pi r_h^2} \int_{\infty_2}^r Y_2(\xi) j_{Li}^{(m)(n)} (\xi) d\xi + \frac{Y_2(r)}{2\i\pi r_h^2} \int_r^{\infty_1} Y_1(\xi) j_{Li}^{(m)(n)}(\xi) d\xi, \nonumber \\
A_{Ri}^{(m)(n)}(r) &= \int_{\infty_2}^{\infty_1} G_Y(r,\xi) j_{Ri}^{(m)(n)} (\xi) d\xi \nonumber \\
& = \frac{Y_1(r)}{2\i\pi r_h^2} \int_{\infty_2}^r Y_2(\xi) j_{Ri}^{(m)(n)} (\xi) d\xi + \frac{Y_2(r)}{2\i\pi r_h^2} \int_r^{\infty_1} Y_1(\xi) j_{Ri}^{(m)(n)}(\xi) d\xi,  \label{A_LRii_mn}
\end{align}
where $Y_1(r)$ and $Y_2(r)$ are two linearly independent solutions for the homogeneous part of \eqref{dynamical_eom_perturb}. Practically, we take them to be \cite{Baggioli:2023tlc,Bu:2024oyz}
\begin{align}
Y_1(r) = -\frac{1}{2} \log \frac{r^2-r_h^2}{r^2 +r_h^2} + \i \pi, \qquad \qquad Y_2(r) = -\frac{1}{2} \log \frac{r^2-r_h^2}{r^2 + r_h^2}.
\end{align}

{\bf Perturbative solutions for the scalar sector}

Owing to the nontrivial mass term in \eqref{5Dbulkaction}, it is impossible to have analytical solutions for the bulk scalar field even perturbatively. While we resort to a numerical technique, we carefully explore the structure of perturbative solutions and reduce the usage of numerical computations to a minimal setting.

First, we consider the linear order solution $X^{(1)}$ in the expansion \eqref{alpha-expansion}. In the Fourier space achieved by $\partial_\mu \to \i k_\mu = (-\i \omega, \i \vec q\,)$, the equation of motion for $X^{(1)}$ is
\begin{align}
\partial_r\left[ r^5 f(r) \partial_r X^{(1)} \right] - \left( m_0^2 - \frac{\mu^2}{r^2} \right) r^3 X^{(1)} - 2 \i \omega r^3 \partial_r X^{(1)} - 3 \i \omega r^2 X^{(1)} - q^2 r X^{(1)} = 0 \label{eom_X1}
\end{align}
Following the idea of \cite{Chakrabarty:2019aeu,Bu:2021jlp,Bu:2022oty}, the solution for $X^{(1)}$ is
\begin{align}
X^{(1)}(r,k_\mu) & = \left[ \frac{1}{2}\coth\left(\frac{\beta\omega}{2} \right) \Sigma_{a}(k_\mu) + \Sigma_{r}(k_\mu) \right] \frac{\Phi(r,k_\mu)}{\Phi^{(3)}(k_\mu)} \nonumber \\
& - \Sigma_{a}(k_\mu) \frac{e^{2\i \omega \chi(r)} } {(1-e^{-\beta\omega})} \frac{\Phi(r,-k_\mu)} {\Phi^{(3)}(-k_\mu)} \label{X1_solution}
\end{align}
Here, $\Phi(r,k_\mu)$ is a regular solution (i.e., the ingoing mode) for the linear equation \eqref{eom_X1}, which will be constructed numerically. Near the AdS boundary $r=\infty$, the regular solution $\Phi(r,k_\mu)$ is expanded as
\begin{align}
\Phi(r\to \infty,k_\mu) = \cdots + \frac{\Phi^{(3)}(k_\mu)}{r^3} + \cdots
\end{align}
The factor $\chi(r)$ in \eqref{X1_solution} is
\begin{align}
\chi(r) \equiv \int_{\infty_2}^r \frac{dy}{y^2 f(y)} = - \frac{1}{4r_{h}} \left[ \pi - 2 \arctan \left(\frac{r}{r_{h}}\right) + \log \left(1+\frac{r_{h}}{r} \right) -\log \left(1- \frac{r_{h}}{r} \right) \right] \label{chi}
\end{align}

Based on the linear solution \eqref{X1_solution}, the higher order solutions can be constructed via the Green's function method as implemented for the gauge sector, see \eqref{A_LRii_mn}. Here, the two linearly independent solutions $\tilde Z_1(r)$ and $\tilde Z_2(r)$ for the homogeneous equation $\Box_X Z(r) = 0$ can be extracted from \eqref{X1_solution}. The result is
\begin{align}
\tilde Z_1(r) = \Phi_0(r), \qquad  \tilde Z_2(r) = \chi(r) \Phi_0(r) - \Phi_1(r)
\end{align}
where $\Phi_0(r)$ and $\Phi_1(r)$ correspond to hydrodynamic expansion of the regular solution $\Phi(r,k_\mu)$
\begin{align}
\Phi(r,k_\mu \to 0) = \Phi_0(r) + \i \omega \Phi_1(r) + \cdots
\end{align}
In practice, we make linear combination of $\tilde Z_1(r)$ and $\tilde Z_2(r)$ and generate two new linear solutions
\begin{align}
Z_1(r) = g_1 \tilde Z_1(r) + g_2 \tilde Z_2(r), \qquad Z_2(r) = h_1 \tilde Z_1(r) + h_2 \tilde Z_2(r) \label{linear_solution_scalar}
\end{align}
which have ``ideal'' asymptotic behaviors near the AdS boundary
\begin{align}
Z_1(r \to \infty_1) = \frac{1}{r} + \cdots + \frac{0}{r^3} + \cdots, \qquad  Z_2(r \to \infty_2) = \frac{0}{r} + \cdots + \frac{1}{r^3} + \cdots
\end{align}

Recall that we will focus on the regime near the phase transition so that we can take \eqref{mu_expansion}. So, throughout the holographic derivation of the EFT action, our computation will be limited to the critical point $\mu = \mu_c$ except for $b_0$ which requires a tiny deviation $\delta\mu$. When $\mu = \mu_c$ (at the critical point), the numerical values for $g_{1,2}, h_{1,2}$ of \eqref{linear_solution_scalar} are\footnote{We have set $r_h =1$ when solving linear solution for the scalar sector. The factor $r_h$ can be easily recovered by dimensional analysis.}
\begin{align}
g_1=-15.04-5.34\text{i},\quad g_2=3.40,\quad h_1=-15.04,\quad h_2=3.40
\end{align}

Now, we present the solutions for the higher order parts of scalar sector
\begin{align}
X^{(m)(n)}(r) & = \int_{\infty_2}^{\infty_1} dr^\prime G_X(r,r^\prime) j_X^{(m)(n)}(r^\prime) dr^\prime \nonumber \\
&= \frac{Z_1(r)}{\mathcal C} \int_{\infty_2}^r Z_2(r') j_X^{(m)(n)} (r') dr' + \frac{Z_2(r)}{\mathcal C} \int_r^{\infty_1} Z_1(r') j_X^{(m)(n)}(r') dr', \label{X_mn}
\end{align}
where $G_X(r,r^\prime)$ is the Green's function. The constant $\mathcal C$ is determined from Wronskian determinant of $Z_1(r)$ and $Z_2(r)$
\begin{align}
W_Z \equiv Z_2(r) \partial_r Z_1(r) - Z_1(r) \partial_r Z_2(r) = \frac{\mathcal C}{r^5f(r)} \Longrightarrow \mathcal C = 18.17 \i
\end{align}

\subsection{Holographic effective action} \label{bdy_action}

In this section, we compute the boundary effective action based on the perturbative solutions obtained in last section \ref{perturb_theory}.

In accord with the expansion of \eqref{alpha-expansion}, the gauge field strength $F_L$ can be expanded as (similarly for $F_R$)
\begin{align}
F_L = \alpha F_L^{(1)} + \alpha^2 F_L^{(2)} + \alpha^3 F_L^{(3)} + \cdots
\end{align}
where for simplicity we ignored both Lorentzian indices and flavor indices. In the bulk action, the contribution from the gauge field strengths is
\begin{align}
S_F & = - \frac{1}{4} \int d^5x \sqrt{-g}\, {\rm Tr}\left(F_L^2 + F_R^2 \right) + S_{ct}^A \nonumber \\
& = - \frac{1}{4} \int d^5x \sqrt{-g} \, {\rm Tr}\left[ \left( F_L^{(1)} \right)^2 + 2 F_L^{(1)} F_L^{(2)} + 2 F_L^{(1)} F_L^{(3)} + \left( F_L^{(2)} \right)^2 + \cdots \right. \nonumber \\
& \qquad \qquad \qquad \qquad \quad \,\, \left. \left( F_R^{(1)} \right)^2 + 2 F_R^{(1)} F_R^{(2)} + 2 F_R^{(1)} F_R^{(3)} + \left( F_R^{(2)} \right)^2 + \cdots \right] + S_{ct}^A \label{S_F}
\end{align}
Then, based on \eqref{S_F}, it can be demonstrated that \cite{Bu:2022esd} the linear solutions $A_L^{(1)}$ and $A_R^{(1)}$ are sufficient in calculating boundary action up to order $\mathcal O(\alpha^4)$\footnote{One may wonder whether the linear scalar solution $X^{(1)}$ will contribute to quartic action through $(F^{(2)}_{L,\, R})^2$. We have carefully checked this and found that the contributions either have higher derivatives or contain more $a$-variables, which we do not cover in section \ref{EFT_constru}.}. Moreover, the terms of order $\mathcal O(\alpha^4)$ in \eqref{S_F} contain at least one boundary derivative, which we have not covered in section \ref{EFT_constru}. Therefore, the contribution from $S_F$ could be simply computed as
\begin{align}
S_F = - \frac{1}{4} \int d^4x \int_{\infty_2}^{\infty_1} dr \sqrt{-g} \left(F_L^2 + F_R^2 \right)\bigg|_{A_L \to A_L^{(1)}, \, A_R \to A_R^{(1)}} + S_{ct}^A \label{S_F_radial_integral}
\end{align}
where the linear solutions $A_L^{(1)}$ and $A_R^{(1)}$ are presented in \eqref{A_LR_00} through \eqref{A_LRii_mn}. Evaluating the radial integral in \eqref{S_F_radial_integral}, we obtain exactly \eqref{L_diff} and the last eight terms of \eqref{L_BCSIGMA_3} with holographic results for various coefficients \cite{Bu:2020jfo,Bu:2024oyz}
\begin{align}
& a_0 = 2r_h^2, \quad a_1 =0, \quad a_2 = -r_h, \quad a_3 = - \frac{\log (2r_h^2/L^2)}{2} , \quad a_4 = - \frac{\log (2r_h^2/L^2)}{2} ,  \nonumber \\
& a_5 = \log (2r_h^2/L^2), \quad a_6=\frac{1}{2}\log (r_h^2/L^2),\quad a_7 =2r_h^2, \quad a_8 =0, \quad a_9 =-r_h , \nonumber \\
&a_{10} = - \frac{\log (2r_h^2/L^2)}{2}, \quad a_{11} = - \frac{\log (2r_h^2/L^2)}{2}, \quad a_{12} = \log (2r_h^2/L^2), \quad a_{13} = \frac{1}{2}\log (r_h^2/L^2), \nonumber\\
&\varpi_{1}=\varpi_{2}=\log (2r_h/L),\qquad \varpi_{3}=\varpi_{4}=-\log(2r_h^2/L^2),\qquad \varpi_{5}=\varpi_{6}=0
\end{align}
where we recovered AdS radius $L$ by dimensional analysis. As pointed out in \cite{Baggioli:2023tlc}, the fact that $a_1=0$ and $a_8 =0$ is related to the hydrodynamic frame that holographic models naturally choose. In other words, $a_1$ and $a_8$ can be consistently set to zero by appropriate field redefinitions, at the cost of having an additional higher order terms \cite{Baggioli:2023tlc,Glorioso:2017fpd}. The results for $a_6$ and $a_{13}$ are renormalization scheme-dependent, see \eqref{Sct_A_X}. Notice that $\varpi_5$ and $\varpi_6$ vanish in the holographic model. 
From the lesson of previous studies \cite{Bu:2022esd,Bu:2024oyz}, this might arise from the probe limit and vanishing of isospin chemical potential. Studies beyond these approximations will be left as future projects to be briefly discussed in section \ref{sum_and_outl}.

We turn to the contribution from scalar sector in the bulk action \eqref{5Dbulkaction}
\begin{align}
S_X & \equiv \int d^5x \sqrt{-g}\, {\rm Tr} \left\{-\left|DX\right|^2 - \left( m_0^2 - \frac{\mu_c^2}{r^2} \right) |X|^2 - a |X|^4 \right\} + S_{ct}^X + S_{bdy} \nonumber \\
&=-\int d^4x \mathrm{Tr} \left( r^{5} X^{\dagger} \partial_{r} X + r^{3} X^{\dagger} \partial_0 X \right) \bigg|_{r=\infty_{2}}^{r=\infty_{1}} +  a \int d^5x \sqrt{-g}\,{\rm Tr}(|X|^4) \nonumber \\
& = \int d^4x\, {\rm Tr}\left( 2 m_1^\dag \Sigma_1 - 2 m_2^\dag \Sigma_2\right) + a \int d^5x \sqrt{-g}\, {\rm Tr} \left( |X|^4 \right) \nonumber \\
& = \int d^4x\, {\rm Tr}\left[ (m_1 + m_2)^\dag \Sigma_a + 2 (m_1 - m_2)^\dag \Sigma_r \right] + a \int d^5x \sqrt{-g}\, {\rm Tr} \left( |X|^4 \right), \label{S_X_radial_integral}
\end{align}
where in the second equality we have integrated by part and made use of scalar's equation of motion. The $m_{1,2}$ are leading terms in the near boundary asymptotic behavior for $X$, see \eqref{AdSboundarycondition}. In accord with the expansions \eqref{alpha-expansion} and \eqref{lambda-expansion}, we expand $m_s$ formally
\begin{align}
m_s & = \alpha\, m_s^{(1)} + \alpha^2 m^{(2)}_s + \alpha^3 m^{(3)} + \cdots, \nonumber \\
m_s^{(l)} & = m_s^{(l)(0)} + \lambda^1 m_s^{(l)(1)} + \lambda^2 m_s^{(l)(2)} + \cdots
\end{align}

From the linear solution $X^{(1)}$ in \eqref{X1_solution}, it is straightforward to read off $m_s^{(1)}$. In the hydrodynamic limit, they are expanded as
\begin{align}
\left( m_1^{(1)} \right)^\dag + \left( m_2^{(1)} \right)^\dag  & = 0.2904(\mu_c - \mu) \Sigma_r^\dag + (0.348 + 0.01 \i) \partial_0\Sigma_r^\dag  - 0.121  \partial_i^2 \Sigma_r^\dag \nonumber \\
& \quad + (0.022+0.100 \i) \partial_0^2 \Sigma_r^\dag + \cdots, \nonumber \\
2\left[\left( m_1^{(1)} \right)^\dag - \left( m_2^{(1)} \right)^\dag\right] & = 0.2904(\mu_c - \mu) \Sigma_a^\dag - (0.348 + 0.01 \i) \partial_0\Sigma_a^\dag  - 0.121  \partial_i^2 \Sigma_a^\dag \nonumber \\
& \quad - (0.022+0.100 \i) \partial_0^2 \Sigma_a^\dag + \cdots. \label{ms_1_hydro}
\end{align}
Then, plugging \eqref{ms_1_hydro} into \eqref{S_X_radial_integral}, we perfectly produce quadratic terms of \eqref{L_SIGMA}. The holographic results for various coefficients of \eqref{L_SIGMA} are (in unit of $r_h=1$)
\begin{align}
b_0 = 0.290(\mu_c - \mu), \quad b_1 = -0.348 - 0.0100 \i, \quad b_2 = -0.121, \quad b_3 = -0.022-0.100 \i ,
\end{align}
where by dimensional analysis $(\mu_c - \mu) \sim (T-T_c)$ with $T_c$ the critical temperature. Here, we see both $b_1$ and $b_3$ are complex, as allowed by the EFT analysis.

We turn to the cubic terms of boundary action, which generically contain both zeroth and first order derivatives. From the holographic formula \eqref{S_X_radial_integral}, this requires to compute $m_1^{(2)(0)}$ and $m_1^{(2)(1)}$. The latter can be extracted from the formal expressions \eqref{X_mn}
\begin{align}
m_1^{(2)(n)} = \frac{1}{\mathcal C} \int_{\infty_2}^{\infty_1} Z_2(r')j_X^{(2)(n)}(r') dr', \qquad
m_2^{(2)(n)} = \frac{1}{\mathcal C} \int_{\infty_2}^{\infty_1} Z_1(r')j_X^{(2)(n)}(r') dr' \label{m12_20_21}
\end{align}
where $n=0,1$. Here, the relevant sources can be read off from the bulk equations
\begin{align}
j_X^{(2)(0)} & = 0, \nonumber \\
j_X^{(2)(1)} & = -2\i r A_{Li}^{(1)(0)}\partial_i X^{(1)(0)} + 2\i r \partial_i X^{(1)(0)} A_{Ri}^{(1)(0)} -\i r \partial_i A_{Li}^{(1)(0)} X^{(1)(0)} \nonumber \\
& \quad + \i r X^{(1)(0)} \partial_i A_{Ri}^{(1)(0)}  \label{j_X_20}
\end{align}
With $A_{L,R}^{(1)(0)}$ presented in \eqref{A_LR_00} and $X^{(1)(0)}$ easily extracted from \eqref{X1_solution}, we work out the radial integrals of \eqref{m12_20_21} numerically:
\begin{align}
\left( m_1^{(2)} \right)^\dag + \left( m_2^{(2)} \right)^\dag  & =0.242\i \left( \partial_i\Sigma_r^{\dag}B_{ri} - C_{ri}\partial_i\Sigma_r^{\dag} \right) + 0.121\i \left( \Sigma_r^{\dag} \partial_iB_{ri} - \partial_iC_{ri}\Sigma_r^{\dag} \right), \nonumber \\
2\left[\left( m_1^{(2)} \right)^\dag - \left( m_2^{(2)} \right)^\dag\right] & = 0.242\i \left( \partial_i\Sigma_r^{\dag}B_{ai}  + \partial_i\Sigma_a^{\dag} B_{ri} - C_{ai}\partial_i\Sigma_r^{\dag}  - C_{ri}\partial_i\Sigma_a^{\dag} \right) \nonumber \\
& \quad +0.121\i\left( \Sigma_r^{\dag}\partial_iB_{ai}+\Sigma_a^{\dag} \partial_iB_{ri} -\partial_iC_{ai}\Sigma_r^{\dag} - \partial_iC_{ri}\Sigma_a^{\dag} \right). \label{ms_2_hydro}
\end{align}

Finally, we compute the quartic terms of boundary action. The holographic formula \eqref{S_X_radial_integral} implies two sources for the quartic terms. The first one corresponds to the bulk part of \eqref{S_X_radial_integral}, which is computed as
\begin{align}
&\int_{\infty_2}^{\infty_1} dr \sqrt{-g}\, |X|^4 \bigg|_{X \to X^{(1)(0)}} \nonumber \\
=& 0.0449 \Sigma_r\Sigma_a^{\dag}\Sigma_r\Sigma_r^{\dag} + 0.0449 \Sigma_r\Sigma_r^{\dag}\Sigma_a\Sigma_r^{\dag} \label{holo_quartic_terms1}
\end{align}
The second source of quartic terms comes from the first part of \eqref{S_X_radial_integral}, which requires to compute $m_s^{(3)(0)}$
\begin{align}
m_1^{(3)(0)} = \frac{1}{\mathcal C} \int_{\infty_2}^{\infty_1} Z_2(r')j_X^{(3)(0)}(r') dr', \qquad
m_2^{(3)(0)} = \frac{1}{\mathcal C} \int_{\infty_2}^{\infty_1} Z_1(r')j_X^{(3)(0)}(r') dr' \label{m12_30}
\end{align}
Here, the relevant source terms are
\begin{align}
j_X^{(3)(0)} & = \frac{r}{f(r)}A_{L0}^2 X + \frac{r}{f(r)}X A_{R0}^2 - \frac{2r}{f(r)} A_{L0}XA_{R0} -r A_{Li} A_{Li} X \nonumber \\
& \quad - r X A_{Ri} A_{Ri} + 2 r A_{Li} X A_{Ri} + 2a X^\dag X X \bigg|_{A_{L,R}\to A_{L, R}^{(1)(0)},\, X\to X^{(1)(0)}}
\end{align}
Working out the radial integrals in \eqref{m12_30}, we have
\begin{align}
\left( m_1^{(2)} \right)^\dag + \left( m_2^{(2)} \right)^\dag  & = 0.0336 \left( \Sigma_r^{\dag} B_{r0} B_{r0} + C_{r0} C_{r0} \Sigma_r^{\dag} - 2 C_{r0} \Sigma_r^{\dag} B_{r0} \right) \nonumber \\
& -0.121 \left( \Sigma_r^{\dag} B_{ri} B_{ri} + C_{ri} C_{ri} \Sigma_r^{\dag} - 2 C_{ri} \Sigma_r^{\dag} B_{ri} \right) \nonumber \\
& -0.0293 a \Sigma_r^{\dag} \Sigma_r \Sigma_r^{\dag} \nonumber \\
2\left[\left( m_1^{(2)} \right)^\dag - \left( m_2^{(2)} \right)^\dag\right] &= 0.0336 \left( \Sigma_a^{\dag} B_{r0} B_{r0} + C_{r0} C_{r0} \Sigma_a^{\dag} - 2 C_{r0} \Sigma_a^{\dag} B_{r0} + 2 \Sigma_r^{\dag} B_{a0} B_{r0} \right. \nonumber \\
& \qquad \qquad \, \, \left. + 2 C_{a0} C_{r0} \Sigma_r^{\dag} - 2 C_{a0} \Sigma_r^{\dag} B_{r0} - 2 C_{r0} \Sigma_r^{\dag} B_{a0} \right)
\nonumber \\
& -0.121 \left( \Sigma_a^{\dag}B_{ri}B_{ri} + C_{ri}C_{ri}\Sigma_a^{\dag} - 2 C_{ri}\Sigma_a^{\dag}B_{ri} + \Sigma_r^{\dag}B_{ri}B_{ai} + \Sigma_r^{\dag}B_{ai}B_{ri} \right. \nonumber\\
& \qquad \qquad \left. + C_{ai} C_{ri} \Sigma_r^{\dag} + C_{ri} C_{ai} \Sigma_r^{\dag} - 2 C_{ai} \Sigma_r^{\dag} B_{ri} - 2 C_{ri} \Sigma_r^{\dag} B_{ai} \right) \nonumber \\
&-0.0293 a \Sigma_r^{\dag} \Sigma_r \Sigma_a^{\dag} \label{ms_3_hydro}
\end{align}

Plugging \eqref{ms_2_hydro}, \eqref{holo_quartic_terms1} and \eqref{ms_3_hydro} into \eqref{S_X_radial_integral}, we read off holographic results for the coefficients in the cubic terms \eqref{L_BCSIGMA_3} and quartic terms \eqref{L_BCSIGMA_4} (in unit of $r_h$)
\begin{align}
& c_0 = c_1 = 0, \quad  d_0= d_1=0, \quad c_2=0.121, \quad d_2=-0.121, \nonumber \\
& c_3=c_3^*=2c_4=2c_4^*=0.0672,\quad  d_3 = d_3^* = 2d_4 = 2d_4^* = 0.0672, \nonumber \\
& \chi_1=0.0156a, \quad \chi_2=-0.134,\quad \chi_3=-0.134, \quad \chi_4=\chi_4^*=-0.0672.
\end{align}

Our holographic results satisfy all the symmetries summarized in section \ref{variable_symmetry}. In particular, owing to the chemical shift symmetry, we see that some quadratic terms, cubic terms and quartic terms (i.e., those terms hidden behind the covariant spatial derivative $\mathcal D_i$ in \eqref{L_SIGMA} and \eqref{L_BCSIGMA_3}) are linked to each other sharing the same coefficients. This is clearly obeyed by our numerical results, as shown in \eqref{ms_1_hydro} (spatial derivative terms), \eqref{ms_2_hydro} and relevant parts (i.e., the second line, the sixth line, and the seventh line) in \eqref{ms_3_hydro}.

Notice that the holographic model gives $c_0 = c_1 = 0$ and $d_0 = d_1 = 0$. This is directly related to the fact that vertices like $X^\dagger XA_{L0}$ and $X^{\dagger} XA_{R0}$ are absent in the bulk theory, as seen from the first line of the formula \eqref{j_X_20}. 
This may be illustrated as that the tree-level Witten diagram (the Left panel of Figure \ref{Witten_diagram_B0_Sigma_Sigma}) vanishes. However, beyond the saddle point approximation undertaken in this work, such terms may be generated through loop effects in the bulk, see the Right panel of Figure \ref{Witten_diagram_B0_Sigma_Sigma}. Basically, the tree-level diagram corresponds to the large $N_c$ limit, while the loop diagram represents a finite $N_c$ correction.

\tikzset{every picture/.style={line width=0.75pt}} 
\begin{figure}[htbp]
\centering
\begin{tikzpicture}[x=0.75pt,y=0.75pt,yscale=-1,xscale=1]


\draw   (95,160.71) .. controls (95,129.94) and (119.94,105) .. (150.71,105) .. controls (181.48,105) and (206.43,129.94) .. (206.43,160.71) .. controls (206.43,191.48) and (181.48,216.43) .. (150.71,216.43) .. controls (119.94,216.43) and (95,191.48) .. (95,160.71) -- cycle ;
\draw    (150.71,105) -- (150.71,160.71) ;
\draw    (150.71,160.71) -- (110,199.33) ;
\draw    (150.71,160.71) -- (190,200) ;
\draw   (264.29,160.71) .. controls (264.29,129.94) and (289.23,105) .. (320,105) .. controls (350.77,105) and (375.71,129.94) .. (375.71,160.71) .. controls (375.71,191.48) and (350.77,216.43) .. (320,216.43) .. controls (289.23,216.43) and (264.29,191.48) .. (264.29,160.71) -- cycle ;
\draw  [color={rgb, 255:red, 19; green, 28; blue, 254 }  ,draw opacity=1 ] (320,146.77) -- (337.2,174.66) -- (302.8,174.66) -- cycle ;
\draw    (320,105) -- (320,146.77) ;
\draw    (302.8,174.66) -- (280.6,199.6) ;
\draw    (337.2,174.66) -- (358.6,200.6) ;
\draw  [draw opacity=0][fill={rgb, 255:red, 208; green, 2; blue, 27 }  ,fill opacity=1 ] (298.8,174.66) .. controls (298.8,172.45) and (300.59,170.66) .. (302.8,170.66) .. controls (305.01,170.66) and (306.8,172.45) .. (306.8,174.66) .. controls (306.8,176.87) and (305.01,178.66) .. (302.8,178.66) .. controls (300.59,178.66) and (298.8,176.87) .. (298.8,174.66) -- cycle ;
\draw  [draw opacity=0][fill={rgb, 255:red, 208; green, 2; blue, 27 }  ,fill opacity=1 ] (333.2,174.66) .. controls (333.2,172.45) and (334.99,170.66) .. (337.2,170.66) .. controls (339.41,170.66) and (341.2,172.45) .. (341.2,174.66) .. controls (341.2,176.87) and (339.41,178.66) .. (337.2,178.66) .. controls (334.99,178.66) and (333.2,176.87) .. (333.2,174.66) -- cycle ;
\draw  [draw opacity=0][fill={rgb, 255:red, 208; green, 2; blue, 27 }  ,fill opacity=1 ] (316,146.77) .. controls (316,144.56) and (317.79,142.77) .. (320,142.77) .. controls (322.21,142.77) and (324,144.56) .. (324,146.77) .. controls (324,148.98) and (322.21,150.77) .. (320,150.77) .. controls (317.79,150.77) and (316,148.98) .. (316,146.77) -- cycle ;

\draw (97,201) node [anchor=north west][inner sep=0.75pt]   [align=left] {$\Sigma ^{\dagger }$};
\draw (192,203) node [anchor=north west][inner sep=0.75pt]   [align=left] {$\Sigma $};
\draw (145,85) node [anchor=north west][inner sep=0.75pt]   [align=left] {$B_0$};
\draw (100,223) node [anchor=north west][inner sep=0.75pt]   [align=left] {$G\left( B_{0} \Sigma \Sigma ^{\dagger }\right) = 0$};
\draw (314,85) node [anchor=north west][inner sep=0.75pt]   [align=left] {$B_{0}$};
\draw (267,201) node [anchor=north west][inner sep=0.75pt]   [align=left] {$\Sigma ^{\dagger }$};
\draw (360.6,203) node [anchor=north west][inner sep=0.75pt]   [align=left] {$\Sigma $};
\draw (270,223) node [anchor=north west][inner sep=0.75pt]   [align=left] {$G\left( B_{0} \Sigma \Sigma ^{\dagger }\right) \neq 0$};

\end{tikzpicture}

\caption{Witten diagrams for $B_0 \Sigma \Sigma^\dag$-terms in the boundary action. 
Left: the Witten diagram at the tree level. It vanishes simply due to the absence of the bulk vertices $X^{\dagger}XA_{L0}$ and $X^{\dagger}XA_{R0}$. Right: the Witten diagram at the one-loop level. The inner lines forming the bulk loop may represent bulk gauge fields, complex scalar field, and bulk gravitons (if beyond the probe limit). }
\label{Witten_diagram_B0_Sigma_Sigma}
\end{figure}

\section{Summary and Outlook}\label{sum_and_outl}

We have constructed a Wilsonian EFT (in a real-time formalism) which is valid for studying the long-wavelength lone-time dynamics of QCD matter near the chiral phase transition. The dynamical variables contain conserved charge densities associated with the chiral symmetry and the order parameter characterising the $\chi$SB. The inclusion of the latter as a dynamical field is crucial as one focuses on critical regime of chiral phase transition. The EFT Lagrangian is stringently constrained by the set of symmetries postulated for hydrodynamic EFT \cite{Crossley:2015evo,Glorioso:2017fpd,Liu:2018lec}. In particular, the dynamical KMS symmetry and the chemical shift symmetry link certain terms in the EFT action. From the EFT action, we have derived a set of stochastic equations for the chiral charge densities and the chiral condensate, which will be useful for numerical simulations. We found that, with higher order terms ignored properly, the set of stochastic equations resemble the model F of Hohenberg-Halperin classification \cite{RevModPhys.49.435}, which was proposed to study dynamical evolution of critical U(1) superfluid system. The EFT approach provides a systematic way of extending phenomenological stochastic models.

By applying the holographic SK technique of \cite{Glorioso:2018mmw}, we have confirmed the EFT construction by deriving the boundary effective action for a modified AdS/QCD model \cite{Chelabi:2015gpc,Chelabi:2015cwn,Chen:2018msc,Cao:2022csq}. The model naturally incorporates spontaneous $\chi$SB and thus allows one to get access into the critical regime of chiral phase transition. Moreover, the holographic study gives valuable information on the parameters in the EFT, whose microscopic theory is strongly coupled and is usually challenging to study with perturbative method. Intriguingly, we find some coefficients (i.e., $c_{0,1}$ and $d_{0,1}$ in \eqref{L_BCSIGMA_3}) are accidentally zero. We attribute this to the saddle point approximation and the probe limit undertaken in this work.

There are several directions that we hope to address in the future. First, it will be interesting to explore physical consequences of higher order terms in \eqref{stochastic_rho_O}, i.e., those beyond the model F in the Hohenberg-Halperin classification, along the line of \cite{Nahrgang:2020yxm}. This might be important in clarifying non-Gaussianity regarding QCD critical point. Second, it would be straightforward to include effect of explicit breaking of chiral symmetry in the spirit of \cite{Hongo:2024brb}, which utilized a spurious symmetry by associating a transformation rule with the mass matrix (as a source for the order parameter $\Sigma$). Presumably, this effect will render the transition into a crossover. Third, the EFT constructed in this work would be useful in understanding phases of nuclear matter at finite temperature and isospin chemical potential, such as pion superfluid phase (i.e., pion condensation). Last but not the least, it will be interesting to consider more realistic AdS/QCD models such as {\color{red}\cite{DeWolfe:2010he,Finazzo:2016mhm,Knaute:2017opk,
Critelli:2017oub,Cai:2022omk,He:2023ado,Zhao:2023gur,Cai:2024eqa,Hippert:2023bel,
Jokela:2024xgz,Chen:2024mmd,Zhu:2025gxo} } that have taken into account latest lattice results, observational constraints, etc. Study along this line is supposed to provide more realistic information for the parameters appearing in the low-energy EFT.

\section*{Acknowledgements}

We would like to thank Matteo Baggioli, Xuanmin Cao, Danning Li, Zhiwei Li and Xiyang Sun for helpful discussions. We are grateful to the anonymous referee for useful suggestions and comments. This work was supported by the National Natural Science Foundation of China (NSFC) under the grant No. 12375044.

\bibliographystyle{utphys}
\bibliography{chiralholoEFT-PRDv2}

\providecommand{\href}[2]{#2}\begingroup\raggedright\begin{thebibliography}{10}

\bibitem{Cuteri:2021ikv}
F.~Cuteri, O.~Philipsen, and A.~Sciarra, ``{On the order of the QCD chiral
  phase transition for different numbers of quark flavours},''
  \href{http://dx.doi.org/10.1007/JHEP11(2021)141}{{\em JHEP} {\bfseries 11}
  (2021) 141}, \href{http://arxiv.org/abs/2107.12739}{{\ttfamily
  arXiv:2107.12739 [hep-lat]}}.

\bibitem{Jaiswal:2020hvk}
A.~Jaiswal {\em et~al.}, ``{Dynamics of QCD matter \textemdash{} current
  status},'' \href{http://dx.doi.org/10.1142/S0218301321300010}{{\em Int. J.
  Mod. Phys. E} {\bfseries 30} no.~02, (2021) 2130001},
  \href{http://arxiv.org/abs/2007.14959}{{\ttfamily arXiv:2007.14959
  [hep-ph]}}.

\bibitem{MUSES:2023hyz}
{\bfseries MUSES} Collaboration, R.~Kumar {\em et~al.}, ``{Theoretical and
  experimental constraints for the equation of state of dense and hot
  matter},'' \href{http://dx.doi.org/10.1007/s41114-024-00049-6}{{\em Living
  Rev. Rel.} {\bfseries 27} no.~1, (2024) 3},
  \href{http://arxiv.org/abs/2303.17021}{{\ttfamily arXiv:2303.17021
  [nucl-th]}}.

\bibitem{Halasz:1998qr}
A.~M. Halasz, A.~D. Jackson, R.~E. Shrock, M.~A. Stephanov, and J.~J.~M.
  Verbaarschot, ``{On the phase diagram of QCD},''
  \href{http://dx.doi.org/10.1103/PhysRevD.58.096007}{{\em Phys. Rev. D}
  {\bfseries 58} (1998) 096007},
  \href{http://arxiv.org/abs/hep-ph/9804290}{{\ttfamily arXiv:hep-ph/9804290}}.

\bibitem{Bzdak:2019pkr}
A.~Bzdak, S.~Esumi, V.~Koch, J.~Liao, M.~Stephanov, and N.~Xu, ``{Mapping the
  Phases of Quantum Chromodynamics with Beam Energy Scan},''
  \href{http://dx.doi.org/10.1016/j.physrep.2020.01.005}{{\em Phys. Rept.}
  {\bfseries 853} (2020) 1--87},
  \href{http://arxiv.org/abs/1906.00936}{{\ttfamily arXiv:1906.00936
  [nucl-th]}}.

\bibitem{Du:2024wjm}
L.~Du, A.~Sorensen, and M.~Stephanov, ``{The QCD phase diagram and Beam Energy
  Scan physics: a theory overview},''
  \href{http://dx.doi.org/10.1142/S021830132430008X}{{\em Int. J. Mod. Phys. E}
  {\bfseries 33} no.~07, (2024) 2430008},
  \href{http://arxiv.org/abs/2402.10183}{{\ttfamily arXiv:2402.10183
  [nucl-th]}}.

\bibitem{Luo:2020pef}
X.~Luo, S.~Shi, N.~Xu, and Y.~Zhang, ``{A Study of the Properties of the QCD
  Phase Diagram in High-Energy Nuclear Collisions},''
  \href{http://dx.doi.org/10.3390/particles3020022}{{\em Particles} {\bfseries
  3} no.~2, (2020) 278--307}, \href{http://arxiv.org/abs/2004.00789}{{\ttfamily
  arXiv:2004.00789 [nucl-ex]}}.

\bibitem{Senger:2021cfo}
P.~Senger, ``{Heavy-Ion Collisions at FAIR-NICA Energies},''
  \href{http://dx.doi.org/10.3390/particles4020020}{{\em Particles} {\bfseries
  4} no.~2, (2021) 214--226}.

\bibitem{Pisarski:1983ms}
R.~D. Pisarski and F.~Wilczek, ``{Remarks on the Chiral Phase Transition in
  Chromodynamics},'' \href{http://dx.doi.org/10.1103/PhysRevD.29.338}{{\em
  Phys. Rev. D} {\bfseries 29} (1984) 338--341}.

\bibitem{Rajagopal:1992qz}
K.~Rajagopal and F.~Wilczek, ``{Static and dynamic critical phenomena at a
  second order QCD phase transition},''
  \href{http://dx.doi.org/10.1016/0550-3213(93)90502-G}{{\em Nucl. Phys. B}
  {\bfseries 399} (1993) 395--425},
  \href{http://arxiv.org/abs/hep-ph/9210253}{{\ttfamily arXiv:hep-ph/9210253}}.

\bibitem{RevModPhys.49.435}
P.~C. Hohenberg and B.~I. Halperin, ``Theory of dynamic critical phenomena,''
  \href{http://dx.doi.org/10.1103/RevModPhys.49.435}{{\em Rev. Mod. Phys.}
  {\bfseries 49} (Jul, 1977) 435--479}.
  \url{https://link.aps.org/doi/10.1103/RevModPhys.49.435}.

\bibitem{Son:2002ci}
D.~T. Son and M.~A. Stephanov, ``{Real time pion propagation in finite
  temperature QCD},'' \href{http://dx.doi.org/10.1103/PhysRevD.66.076011}{{\em
  Phys. Rev. D} {\bfseries 66} (2002) 076011},
  \href{http://arxiv.org/abs/hep-ph/0204226}{{\ttfamily arXiv:hep-ph/0204226}}.

\bibitem{Son:2001ff}
D.~T. Son and M.~A. Stephanov, ``{Pion propagation near the QCD chiral phase
  transition},'' \href{http://dx.doi.org/10.1103/PhysRevLett.88.202302}{{\em
  Phys. Rev. Lett.} {\bfseries 88} (2002) 202302},
  \href{http://arxiv.org/abs/hep-ph/0111100}{{\ttfamily arXiv:hep-ph/0111100}}.

\bibitem{Son:1999pa}
D.~T. Son, ``{Hydrodynamics of nuclear matter in the chiral limit},''
  \href{http://dx.doi.org/10.1103/PhysRevLett.84.3771}{{\em Phys. Rev. Lett.}
  {\bfseries 84} (2000) 3771--3774},
  \href{http://arxiv.org/abs/hep-ph/9912267}{{\ttfamily arXiv:hep-ph/9912267}}.

\bibitem{Grossi:2020ezz}
E.~Grossi, A.~Soloviev, D.~Teaney, and F.~Yan, ``{Transport and hydrodynamics
  in the chiral limit},''
  \href{http://dx.doi.org/10.1103/PhysRevD.102.014042}{{\em Phys. Rev. D}
  {\bfseries 102} no.~1, (2020) 014042},
  \href{http://arxiv.org/abs/2005.02885}{{\ttfamily arXiv:2005.02885
  [hep-th]}}.

\bibitem{Grossi:2021gqi}
E.~Grossi, A.~Soloviev, D.~Teaney, and F.~Yan, ``{Soft pions and transport near
  the chiral critical point},''
  \href{http://dx.doi.org/10.1103/PhysRevD.104.034025}{{\em Phys. Rev. D}
  {\bfseries 104} no.~3, (2021) 034025},
  \href{http://arxiv.org/abs/2101.10847}{{\ttfamily arXiv:2101.10847
  [nucl-th]}}.

\bibitem{Florio:2021jlx}
A.~Florio, E.~Grossi, A.~Soloviev, and D.~Teaney, ``{Dynamics of the $O(4)$
  critical point in QCD},''
  \href{http://dx.doi.org/10.1103/PhysRevD.105.054512}{{\em Phys. Rev. D}
  {\bfseries 105} no.~5, (2022) 054512},
  \href{http://arxiv.org/abs/2111.03640}{{\ttfamily arXiv:2111.03640
  [hep-lat]}}.

\bibitem{Cao:2022csq}
X.~Cao, M.~Baggioli, H.~Liu, and D.~Li, ``{Pion dynamics in a soft-wall AdS-QCD
  model},'' \href{http://dx.doi.org/10.1007/JHEP12(2022)113}{{\em JHEP}
  {\bfseries 12} (2022) 113}, \href{http://arxiv.org/abs/2210.09088}{{\ttfamily
  arXiv:2210.09088 [hep-ph]}}.

\bibitem{Braun:2023qak}
J.~Braun {\em et~al.}, ``{Soft modes in hot QCD matter},''
  \href{http://arxiv.org/abs/2310.19853}{{\ttfamily arXiv:2310.19853
  [hep-ph]}}.

\bibitem{Roth:2024rbi}
J.~V. Roth, Y.~Ye, S.~Schlichting, and L.~von Smekal, ``{Dynamic critical
  behavior of the chiral phase transition from the real-time functional
  renormalization group},'' \href{http://arxiv.org/abs/2403.04573}{{\ttfamily
  arXiv:2403.04573 [hep-ph]}}.

\bibitem{Crossley:2015evo}
M.~Crossley, P.~Glorioso, and H.~Liu, ``{Effective field theory of dissipative
  fluids},'' \href{http://dx.doi.org/10.1007/JHEP09(2017)095}{{\em JHEP}
  {\bfseries 09} (2017) 095}, \href{http://arxiv.org/abs/1511.03646}{{\ttfamily
  arXiv:1511.03646 [hep-th]}}.

\bibitem{Glorioso:2017fpd}
P.~Glorioso, M.~Crossley, and H.~Liu, ``{Effective field theory of dissipative
  fluids (II): classical limit, dynamical KMS symmetry and entropy current},''
  \href{http://dx.doi.org/10.1007/JHEP09(2017)096}{{\em JHEP} {\bfseries 09}
  (2017) 096}, \href{http://arxiv.org/abs/1701.07817}{{\ttfamily
  arXiv:1701.07817 [hep-th]}}.

\bibitem{Haehl:2015uoc}
F.~M. Haehl, R.~Loganayagam, and M.~Rangamani, ``{Topological sigma models \&
  dissipative hydrodynamics},''
  \href{http://dx.doi.org/10.1007/JHEP04(2016)039}{{\em JHEP} {\bfseries 04}
  (2016) 039}, \href{http://arxiv.org/abs/1511.07809}{{\ttfamily
  arXiv:1511.07809 [hep-th]}}.

\bibitem{Haehl:2018lcu}
F.~M. Haehl, R.~Loganayagam, and M.~Rangamani, ``{Effective Action for
  Relativistic Hydrodynamics: Fluctuations, Dissipation, and Entropy Inflow},''
  \href{http://dx.doi.org/10.1007/JHEP10(2018)194}{{\em JHEP} {\bfseries 10}
  (2018) 194}, \href{http://arxiv.org/abs/1803.11155}{{\ttfamily
  arXiv:1803.11155 [hep-th]}}.

\bibitem{Liu:2018lec}
H.~Liu and P.~Glorioso, ``{Lectures on non-equilibrium effective field theories
  and fluctuating hydrodynamics},''
  \href{http://dx.doi.org/10.22323/1.305.0008}{{\em PoS} {\bfseries 305} (2018)
  008}, \href{http://arxiv.org/abs/1805.09331}{{\ttfamily arXiv:1805.09331
  [hep-th]}}.

\bibitem{Maldacena:1997re}
J.~M. Maldacena, ``{The Large N limit of superconformal field theories and
  supergravity},'' \href{http://dx.doi.org/10.1023/A:1026654312961}{{\em Adv.
  Theor. Math. Phys.} {\bfseries 2} (1998) 231--252},
  \href{http://arxiv.org/abs/hep-th/9711200}{{\ttfamily arXiv:hep-th/9711200}}.

\bibitem{Gubser:1998bc}
S.~S. Gubser, I.~R. Klebanov, and A.~M. Polyakov, ``{Gauge theory correlators
  from noncritical string theory},''
  \href{http://dx.doi.org/10.1016/S0370-2693(98)00377-3}{{\em Phys. Lett. B}
  {\bfseries 428} (1998) 105--114},
  \href{http://arxiv.org/abs/hep-th/9802109}{{\ttfamily arXiv:hep-th/9802109}}.

\bibitem{Witten:1998qj}
E.~Witten, ``{Anti-de Sitter space and holography},''
  \href{http://dx.doi.org/10.4310/ATMP.1998.v2.n2.a2}{{\em Adv. Theor. Math.
  Phys.} {\bfseries 2} (1998) 253--291},
  \href{http://arxiv.org/abs/hep-th/9802150}{{\ttfamily arXiv:hep-th/9802150}}.

\bibitem{Herzog:2002pc}
C.~P. Herzog and D.~T. Son, ``{Schwinger-Keldysh propagators from AdS/CFT
  correspondence},''
  \href{http://dx.doi.org/10.1088/1126-6708/2003/03/046}{{\em JHEP} {\bfseries
  03} (2003) 046}, \href{http://arxiv.org/abs/hep-th/0212072}{{\ttfamily
  arXiv:hep-th/0212072}}.

\bibitem{Skenderis:2008dh}
K.~Skenderis and B.~C. van Rees, ``{Real-time gauge/gravity duality},''
  \href{http://dx.doi.org/10.1103/PhysRevLett.101.081601}{{\em Phys. Rev.
  Lett.} {\bfseries 101} (2008) 081601},
  \href{http://arxiv.org/abs/0805.0150}{{\ttfamily arXiv:0805.0150 [hep-th]}}.

\bibitem{Skenderis:2008dg}
K.~Skenderis and B.~C. van Rees, ``{Real-time gauge/gravity duality:
  Prescription, Renormalization and Examples},''
  \href{http://dx.doi.org/10.1088/1126-6708/2009/05/085}{{\em JHEP} {\bfseries
  05} (2009) 085}, \href{http://arxiv.org/abs/0812.2909}{{\ttfamily
  arXiv:0812.2909 [hep-th]}}.

\bibitem{Glorioso:2018mmw}
P.~Glorioso, M.~Crossley, and H.~Liu, ``{A prescription for holographic
  Schwinger-Keldysh contour in non-equilibrium systems},''
  \href{http://arxiv.org/abs/1812.08785}{{\ttfamily arXiv:1812.08785
  [hep-th]}}.

\bibitem{deBoer:2018qqm}
J.~de~Boer, M.~P. Heller, and N.~Pinzani-Fokeeva, ``{Holographic
  Schwinger-Keldysh effective field theories},''
  \href{http://dx.doi.org/10.1007/JHEP05(2019)188}{{\em JHEP} {\bfseries 05}
  (2019) 188}, \href{http://arxiv.org/abs/1812.06093}{{\ttfamily
  arXiv:1812.06093 [hep-th]}}.

\bibitem{Chakrabarty:2019aeu}
B.~Chakrabarty, J.~Chakravarty, S.~Chaudhuri, C.~Jana, R.~Loganayagam, and
  A.~Sivakumar, ``{Nonlinear Langevin dynamics via holography},''
  \href{http://dx.doi.org/10.1007/JHEP01(2020)165}{{\em JHEP} {\bfseries 01}
  (2020) 165}, \href{http://arxiv.org/abs/1906.07762}{{\ttfamily
  arXiv:1906.07762 [hep-th]}}.

\bibitem{Bu:2020jfo}
Y.~Bu, T.~Demircik, and M.~Lublinsky, ``{All order effective action for charge
  diffusion from Schwinger-Keldysh holography},''
  \href{http://dx.doi.org/10.1007/JHEP05(2021)187}{{\em JHEP} {\bfseries 05}
  (2021) 187}, \href{http://arxiv.org/abs/2012.08362}{{\ttfamily
  arXiv:2012.08362 [hep-th]}}.

\bibitem{Bu:2021clf}
Y.~Bu, M.~Fujita, and S.~Lin, ``{Ginzburg-Landau effective action for a
  fluctuating holographic superconductor},''
  \href{http://dx.doi.org/10.1007/JHEP09(2021)168}{{\em JHEP} {\bfseries 09}
  (2021) 168}, \href{http://arxiv.org/abs/2106.00556}{{\ttfamily
  arXiv:2106.00556 [hep-th]}}.

\bibitem{Bu:2021jlp}
Y.~Bu and B.~Zhang, ``{Schwinger-Keldysh effective action for a relativistic
  Brownian particle in the AdS/CFT correspondence},''
  \href{http://dx.doi.org/10.1103/PhysRevD.104.086002}{{\em Phys. Rev. D}
  {\bfseries 104} no.~8, (2021) 086002},
  \href{http://arxiv.org/abs/2108.10060}{{\ttfamily arXiv:2108.10060
  [hep-th]}}.

\bibitem{Bu:2022esd}
Y.~Bu, X.~Sun, and B.~Zhang, ``{Holographic Schwinger-Keldysh field theory of
  SU(2) diffusion},'' \href{http://dx.doi.org/10.1007/JHEP08(2022)223}{{\em
  JHEP} {\bfseries 08} (2022) 223},
  \href{http://arxiv.org/abs/2205.00195}{{\ttfamily arXiv:2205.00195
  [hep-th]}}.

\bibitem{Bu:2022oty}
Y.~Bu, B.~Zhang, and J.~Zhang, ``{Nonlinear effective dynamics of a Brownian
  particle in magnetized plasma},''
  \href{http://dx.doi.org/10.1103/PhysRevD.106.086014}{{\em Phys. Rev. D}
  {\bfseries 106} no.~8, (2022) 086014},
  \href{http://arxiv.org/abs/2210.02274}{{\ttfamily arXiv:2210.02274
  [hep-th]}}.

\bibitem{Baggioli:2023tlc}
M.~Baggioli, Y.~Bu, and V.~Ziogas, ``{U(1) quasi-hydrodynamics:
  Schwinger-Keldysh effective field theory and holography},''
  \href{http://dx.doi.org/10.1007/JHEP09(2023)019}{{\em JHEP} {\bfseries 09}
  (2023) 019}, \href{http://arxiv.org/abs/2304.14173}{{\ttfamily
  arXiv:2304.14173 [hep-th]}}.

\bibitem{Bu:2024oyz}
Y.~Bu, H.~Gao, X.~Gao, and Z.~Li, ``{Nearly critical superfluid: effective
  field theory and holography},''
  \href{http://dx.doi.org/10.1007/JHEP07(2024)104}{{\em JHEP} {\bfseries 07}
  (2024) 104}, \href{http://arxiv.org/abs/2401.12294}{{\ttfamily
  arXiv:2401.12294 [hep-th]}}.

\bibitem{Liu:2024tqe}
Y.~Liu, Y.-W. Sun, and X.-M. Wu, ``{Holographic Schwinger-Keldysh effective
  field theories including a non-hydrodynamic mode},''
  \href{http://arxiv.org/abs/2411.16306}{{\ttfamily arXiv:2411.16306
  [hep-th]}}.

\bibitem{Baggioli:2024zfq}
M.~Baggioli, Y.~Bu, and X.~Sun, ``{Chiral Anomalous Magnetohydrodynamics in
  action: effective field theory and holography},''
  \href{http://arxiv.org/abs/2412.02361}{{\ttfamily arXiv:2412.02361
  [hep-th]}}.

\bibitem{Ho:2013rra}
S.-H. Ho, W.~Li, F.-L. Lin, and B.~Ning, ``{Quantum Decoherence with
  Holography},'' \href{http://dx.doi.org/10.1007/JHEP01(2014)170}{{\em JHEP}
  {\bfseries 01} (2014) 170}, \href{http://arxiv.org/abs/1309.5855}{{\ttfamily
  arXiv:1309.5855 [hep-th]}}.

\bibitem{Ghosh:2020lel}
J.~K. Ghosh, R.~Loganayagam, S.~G. Prabhu, M.~Rangamani, A.~Sivakumar, and
  V.~Vishal, ``{Effective field theory of stochastic diffusion from gravity},''
  \href{http://dx.doi.org/10.1007/JHEP05(2021)130}{{\em JHEP} {\bfseries 05}
  (2021) 130}, \href{http://arxiv.org/abs/2012.03999}{{\ttfamily
  arXiv:2012.03999 [hep-th]}}.

\bibitem{He:2021jna}
T.~He, R.~Loganayagam, M.~Rangamani, and J.~Virrueta, ``{An effective
  description of momentum diffusion in a charged plasma from holography},''
  \href{http://arxiv.org/abs/2108.03244}{{\ttfamily arXiv:2108.03244
  [hep-th]}}.

\bibitem{Son:2004iv}
D.~T. Son and M.~A. Stephanov, ``{Dynamic universality class of the QCD
  critical point},'' \href{http://dx.doi.org/10.1103/PhysRevD.70.056001}{{\em
  Phys. Rev. D} {\bfseries 70} (2004) 056001},
  \href{http://arxiv.org/abs/hep-ph/0401052}{{\ttfamily arXiv:hep-ph/0401052}}.

\bibitem{Glorioso:2020loc}
P.~Glorioso, L.~V. Delacr\'etaz, X.~Chen, R.~M. Nandkishore, and A.~Lucas,
  ``{Hydrodynamics in lattice models with continuous non-Abelian symmetries},''
  \href{http://dx.doi.org/10.21468/SciPostPhys.10.1.015}{{\em SciPost Phys.}
  {\bfseries 10} no.~1, (2021) 015},
  \href{http://arxiv.org/abs/2007.13753}{{\ttfamily arXiv:2007.13753
  [cond-mat.stat-mech]}}.

\bibitem{Hongo:2024brb}
M.~Hongo, N.~Sogabe, M.~A. Stephanov, and H.-U. Yee, ``{Schwinger-Keldysh
  effective action for hydrodynamics with approximate symmetries},''
  \href{http://arxiv.org/abs/2411.08016}{{\ttfamily arXiv:2411.08016
  [hep-th]}}.

\bibitem{Donos:2023ibv}
A.~Donos and P.~Kailidis, ``{Nearly critical superfluids in Keldysh-Schwinger
  formalism},'' \href{http://dx.doi.org/10.1007/JHEP01(2024)110}{{\em JHEP}
  {\bfseries 01} (2024) 110}, \href{http://arxiv.org/abs/2304.06008}{{\ttfamily
  arXiv:2304.06008 [hep-th]}}.

\bibitem{Natsuume:2010bs}
M.~Natsuume and T.~Okamura, ``{Dynamic universality class of large-N gauge
  theories},'' \href{http://dx.doi.org/10.1103/PhysRevD.83.046008}{{\em Phys.
  Rev. D} {\bfseries 83} (2011) 046008},
  \href{http://arxiv.org/abs/1012.0575}{{\ttfamily arXiv:1012.0575 [hep-th]}}.

\bibitem{Chelabi:2015gpc}
K.~Chelabi, Z.~Fang, M.~Huang, D.~Li, and Y.-L. Wu, ``{Chiral Phase Transition
  in the Soft-Wall Model of AdS/QCD},''
  \href{http://dx.doi.org/10.1007/JHEP04(2016)036}{{\em JHEP} {\bfseries 04}
  (2016) 036}, \href{http://arxiv.org/abs/1512.06493}{{\ttfamily
  arXiv:1512.06493 [hep-ph]}}.

\bibitem{Chelabi:2015cwn}
K.~Chelabi, Z.~Fang, M.~Huang, D.~Li, and Y.-L. Wu, ``{Realization of chiral
  symmetry breaking and restoration in holographic QCD},''
  \href{http://dx.doi.org/10.1103/PhysRevD.93.101901}{{\em Phys. Rev. D}
  {\bfseries 93} no.~10, (2016) 101901},
  \href{http://arxiv.org/abs/1511.02721}{{\ttfamily arXiv:1511.02721
  [hep-ph]}}.

\bibitem{Chen:2018msc}
J.~Chen, S.~He, M.~Huang, and D.~Li, ``{Critical exponents of finite
  temperature chiral phase transition in soft-wall AdS/QCD models},''
  \href{http://dx.doi.org/10.1007/JHEP01(2019)165}{{\em JHEP} {\bfseries 01}
  (2019) 165}, \href{http://arxiv.org/abs/1810.07019}{{\ttfamily
  arXiv:1810.07019 [hep-ph]}}.

\bibitem{Erlich:2005qh}
J.~Erlich, E.~Katz, D.~T. Son, and M.~A. Stephanov, ``{QCD and a holographic
  model of hadrons},''
  \href{http://dx.doi.org/10.1103/PhysRevLett.95.261602}{{\em Phys. Rev. Lett.}
  {\bfseries 95} (2005) 261602},
  \href{http://arxiv.org/abs/hep-ph/0501128}{{\ttfamily arXiv:hep-ph/0501128}}.

\bibitem{Kardar:1986xt}
M.~Kardar, G.~Parisi, and Y.-C. Zhang, ``{Dynamic Scaling of Growing
  Interfaces},'' \href{http://dx.doi.org/10.1103/PhysRevLett.56.889}{{\em Phys.
  Rev. Lett.} {\bfseries 56} (1986) 889}.

\bibitem{Donos:2022qao}
A.~Donos and P.~Kailidis, ``{Nearly critical holographic superfluids},''
  \href{http://dx.doi.org/10.1007/JHEP12(2022)028}{{\em JHEP} {\bfseries 12}
  (2022) 028}, \href{http://arxiv.org/abs/2210.06513}{{\ttfamily
  arXiv:2210.06513 [hep-th]}}. [Erratum: JHEP 07, 232 (2023)].

\bibitem{Gubser:2008ny}
S.~S. Gubser and A.~Nellore, ``{Mimicking the QCD equation of state with a dual
  black hole},'' \href{http://dx.doi.org/10.1103/PhysRevD.78.086007}{{\em Phys.
  Rev. D} {\bfseries 78} (2008) 086007},
  \href{http://arxiv.org/abs/0804.0434}{{\ttfamily arXiv:0804.0434 [hep-th]}}.

\bibitem{Gubser:2008yx}
S.~S. Gubser, A.~Nellore, S.~S. Pufu, and F.~D. Rocha, ``{Thermodynamics and
  bulk viscosity of approximate black hole duals to finite temperature quantum
  chromodynamics},''
  \href{http://dx.doi.org/10.1103/PhysRevLett.101.131601}{{\em Phys. Rev.
  Lett.} {\bfseries 101} (2008) 131601},
  \href{http://arxiv.org/abs/0804.1950}{{\ttfamily arXiv:0804.1950 [hep-th]}}.

\bibitem{Gursoy:2008bu}
U.~Gursoy, E.~Kiritsis, L.~Mazzanti, and F.~Nitti, ``{Deconfinement and Gluon
  Plasma Dynamics in Improved Holographic QCD},''
  \href{http://dx.doi.org/10.1103/PhysRevLett.101.181601}{{\em Phys. Rev.
  Lett.} {\bfseries 101} (2008) 181601},
  \href{http://arxiv.org/abs/0804.0899}{{\ttfamily arXiv:0804.0899 [hep-th]}}.

\bibitem{Gursoy:2008za}
U.~Gursoy, E.~Kiritsis, L.~Mazzanti, and F.~Nitti, ``{Holography and
  Thermodynamics of 5D Dilaton-gravity},''
  \href{http://dx.doi.org/10.1088/1126-6708/2009/05/033}{{\em JHEP} {\bfseries
  05} (2009) 033}, \href{http://arxiv.org/abs/0812.0792}{{\ttfamily
  arXiv:0812.0792 [hep-th]}}.

\bibitem{DeWolfe:2010he}
O.~DeWolfe, S.~S. Gubser, and C.~Rosen, ``{A holographic critical point},''
  \href{http://dx.doi.org/10.1103/PhysRevD.83.086005}{{\em Phys. Rev. D}
  {\bfseries 83} (2011) 086005},
  \href{http://arxiv.org/abs/1012.1864}{{\ttfamily arXiv:1012.1864 [hep-th]}}.

\bibitem{Finazzo:2016mhm}
S.~I. Finazzo, R.~Critelli, R.~Rougemont, and J.~Noronha, ``{Momentum transport
  in strongly coupled anisotropic plasmas in the presence of strong magnetic
  fields},'' \href{http://dx.doi.org/10.1103/PhysRevD.94.054020}{{\em Phys.
  Rev. D} {\bfseries 94} no.~5, (2016) 054020},
  \href{http://arxiv.org/abs/1605.06061}{{\ttfamily arXiv:1605.06061
  [hep-ph]}}. [Erratum: Phys.Rev.D 96, 019903 (2017)].

\bibitem{Knaute:2017opk}
J.~Knaute, R.~Yaresko, and B.~K\"ampfer, ``{Holographic QCD phase diagram with
  critical point from Einstein\textendash{}Maxwell-dilaton dynamics},''
  \href{http://dx.doi.org/10.1016/j.physletb.2018.01.053}{{\em Phys. Lett. B}
  {\bfseries 778} (2018) 419--425},
  \href{http://arxiv.org/abs/1702.06731}{{\ttfamily arXiv:1702.06731
  [hep-ph]}}.

\bibitem{Critelli:2017oub}
R.~Critelli, J.~Noronha, J.~Noronha-Hostler, I.~Portillo, C.~Ratti, and
  R.~Rougemont, ``{Critical point in the phase diagram of primordial
  quark-gluon matter from black hole physics},''
  \href{http://dx.doi.org/10.1103/PhysRevD.96.096026}{{\em Phys. Rev. D}
  {\bfseries 96} no.~9, (2017) 096026},
  \href{http://arxiv.org/abs/1706.00455}{{\ttfamily arXiv:1706.00455
  [nucl-th]}}.

\bibitem{Crossley:2015tka}
M.~Crossley, P.~Glorioso, H.~Liu, and Y.~Wang, ``{Off-shell hydrodynamics from
  holography},'' \href{http://dx.doi.org/10.1007/JHEP02(2016)124}{{\em JHEP}
  {\bfseries 02} (2016) 124}, \href{http://arxiv.org/abs/1504.07611}{{\ttfamily
  arXiv:1504.07611 [hep-th]}}.

\bibitem{Skenderis:2002wp}
K.~Skenderis, ``{Lecture notes on holographic renormalization},''
  \href{http://dx.doi.org/10.1088/0264-9381/19/22/306}{{\em Class. Quant.
  Grav.} {\bfseries 19} (2002) 5849--5876},
  \href{http://arxiv.org/abs/hep-th/0209067}{{\ttfamily arXiv:hep-th/0209067}}.

\bibitem{Round:2010kj}
M.~Round, ``{Holographic Renormalisation and the Electroweak Precision
  Parameters},'' \href{http://dx.doi.org/10.1103/PhysRevD.82.053002}{{\em Phys.
  Rev. D} {\bfseries 82} (2010) 053002},
  \href{http://arxiv.org/abs/1003.2933}{{\ttfamily arXiv:1003.2933 [hep-ph]}}.

\bibitem{Marolf:2006nd}
D.~Marolf and S.~F. Ross, ``{Boundary Conditions and New Dualities: Vector
  Fields in AdS/CFT},''
  \href{http://dx.doi.org/10.1088/1126-6708/2006/11/085}{{\em JHEP} {\bfseries
  11} (2006) 085}, \href{http://arxiv.org/abs/hep-th/0606113}{{\ttfamily
  arXiv:hep-th/0606113}}.

\bibitem{Nahrgang:2020yxm}
M.~Nahrgang and M.~Bluhm, ``{Modeling the diffusive dynamics of critical
  fluctuations near the QCD critical point},''
  \href{http://dx.doi.org/10.1103/PhysRevD.102.094017}{{\em Phys. Rev. D}
  {\bfseries 102} no.~9, (2020) 094017},
  \href{http://arxiv.org/abs/2007.10371}{{\ttfamily arXiv:2007.10371
  [nucl-th]}}.

\bibitem{Cai:2022omk}
R.-G. Cai, S.~He, L.~Li, and Y.-X. Wang, ``{Probing QCD critical point and
  induced gravitational wave by black hole physics},''
  \href{http://dx.doi.org/10.1103/PhysRevD.106.L121902}{{\em Phys. Rev. D}
  {\bfseries 106} no.~12, (2022) L121902},
  \href{http://arxiv.org/abs/2201.02004}{{\ttfamily arXiv:2201.02004
  [hep-th]}}.

\bibitem{He:2023ado}
S.~He, L.~Li, S.~Wang, and S.-J. Wang, ``{Constraints on holographic QCD phase
  transitions from PTA observations},''
  \href{http://dx.doi.org/10.1007/s11433-024-2468-x}{{\em Sci. China Phys.
  Mech. Astron.} {\bfseries 68} no.~1, (2025) 210411},
  \href{http://arxiv.org/abs/2308.07257}{{\ttfamily arXiv:2308.07257
  [hep-ph]}}.

\bibitem{Zhao:2023gur}
Y.-Q. Zhao, S.~He, D.~Hou, L.~Li, and Z.~Li, ``{Phase structure and critical
  phenomena in two-flavor QCD by holography},''
  \href{http://dx.doi.org/10.1103/PhysRevD.109.086015}{{\em Phys. Rev. D}
  {\bfseries 109} no.~8, (2024) 086015},
  \href{http://arxiv.org/abs/2310.13432}{{\ttfamily arXiv:2310.13432
  [hep-ph]}}.

\bibitem{Cai:2024eqa}
R.-G. Cai, S.~He, L.~Li, and H.-A. Zeng, ``{QCD Phase Diagram at finite
  Magnetic Field and Chemical Potential: A Holographic Approach Using Machine
  Learning},'' \href{http://arxiv.org/abs/2406.12772}{{\ttfamily
  arXiv:2406.12772 [hep-th]}}.

\bibitem{Hippert:2023bel}
M.~Hippert, J.~Grefa, T.~A. Manning, J.~Noronha, J.~Noronha-Hostler,
  I.~Portillo~Vazquez, C.~Ratti, R.~Rougemont, and M.~Trujillo, ``{Bayesian
  location of the QCD critical point from a holographic perspective},''
  \href{http://dx.doi.org/10.1103/PhysRevD.110.094006}{{\em Phys. Rev. D}
  {\bfseries 110} no.~9, (2024) 094006},
  \href{http://arxiv.org/abs/2309.00579}{{\ttfamily arXiv:2309.00579
  [nucl-th]}}.

\bibitem{Jokela:2024xgz}
N.~Jokela, M.~J\"arvinen, and A.~Piispa, ``{Refining holographic models of the
  quark-gluon plasma},''
  \href{http://dx.doi.org/10.1103/PhysRevD.110.126013}{{\em Phys. Rev. D}
  {\bfseries 110} no.~12, (2024) 126013},
  \href{http://arxiv.org/abs/2405.02394}{{\ttfamily arXiv:2405.02394
  [hep-th]}}.

\bibitem{Chen:2024mmd}
X.~Chen and M.~Huang, ``{Flavor dependent critical endpoint from holographic
  QCD through machine learning},''
  \href{http://dx.doi.org/10.1007/JHEP02(2025)123}{{\em JHEP} {\bfseries 02}
  (2025) 123}, \href{http://arxiv.org/abs/2405.06179}{{\ttfamily
  arXiv:2405.06179 [hep-ph]}}.

\bibitem{Zhu:2025gxo}
L.~Zhu, X.~Chen, K.~Zhou, H.~Zhang, and M.~Huang, ``{Bayesian Inference of the
  Critical Endpoint in 2+1-Flavor System from Holographic QCD},''
  \href{http://arxiv.org/abs/2501.17763}{{\ttfamily arXiv:2501.17763
  [hep-ph]}}.

\end{thebibliography}\endgroup
\end{document}